\documentclass[12pt]{article}
\usepackage[T1]{fontenc}
\usepackage{kpfonts}
\usepackage[dvipsnames]{xcolor}
\usepackage{pdfpages}
\usepackage{sgame}
\usepackage{accents}
\usepackage{tikz-cd}
\usepackage{float}
\usepackage[round]{natbib}   
\usepackage{multirow}
\usepackage{dutchcal}
\usepackage{pdfpages}
\usepackage{bbm}
\usepackage{sgame}
\usepackage{mathtools}
\usepackage{amsthm}
\usepackage{enumitem}
\usepackage[margin=1in]{geometry}
\usepackage{caption}
\usepackage{subcaption}
\usepackage{epigraph}
\usepackage{listings}
\usetikzlibrary{calc}
\usetikzlibrary{shapes,arrows}

\DeclareMathOperator\supp{supp}
\DeclareMathOperator\inter{int}

\DeclareMathOperator*{\argmax}{arg\,max}
\DeclareMathOperator*{\argmin}{arg\,min}

\newtheorem{theorem}{Theorem}[section]
\newtheorem{proposition}[theorem]{Proposition}
\newtheorem{lemma}[theorem]{Lemma}
\newtheorem{corollary}[theorem]{Corollary}

\newtheorem{observation}[theorem]{Observation}

\theoremstyle{definition}

\newtheorem{remark}[theorem]{Remark}

\usepackage{hyperref}
\hypersetup{
    colorlinks=true,
    linkcolor=CadetBlue,
    filecolor=CadetBlue,      
    urlcolor=CadetBlue,
    citecolor=CadetBlue,
}

\definecolor{backcolour}{rgb}{0.63, 0.79, 0.95}

\lstdefinestyle{mystyle}{
  backgroundcolor=\color{backcolour},
  basicstyle=\ttfamily\footnotesize,
  breakatwhitespace=false,         
  breaklines=true,                 
  captionpos=b,                    
  keepspaces=true,                 
  numbers=left,                    
  numbersep=5pt,                  
  showspaces=false,                
  showstringspaces=false,
  showtabs=false,                  
  tabsize=2
}

\providecommand{\keywords}[1]{\textbf{\textit{Keywords:}} #1}
\providecommand{\jel}[1]{\textbf{\textit{JEL Classifications:}} #1}

\begin{document}
\author{Mark Whitmeyer\thanks{Arizona State University, Email: \href{mailto:mark.whitmeyer@gmail.com}{mark.whitmeyer@gmail.com}} \and Kun Zhang\thanks{Arizona State University, Email: \href{mailto:kunzhang@asu.edu}{kunzhang@asu.edu} \newline We thank Brian Albrecht, Hector Chade, Vasudha Jain, Andreas Kleiner, Alejandro Manelli, Teemu Pekkarinen, Ludvig Sinander, Eddie Schlee, Ina Taneva, Can Urgun, Han Wang, Joseph Whitmeyer, Thomas Wiseman, and Hanzhe Zhang for their advice. We also appreciate the useful feedback from conference audiences at AMES, ESEM, MWET, NASMES, and Stony Brook 2022; and seminar audiences at ASU, Arizona, CUHK-HKU-HKUST, Fordham, Princeton, and Warwick.}}

\title{Buying Opinions}

\date{\today}

\maketitle

\begin{abstract}A principal hires an agent to acquire soft information about an unknown state. Even though neither \textit{how} the agent learns nor \textit{what} the agent discovers are contractible, we show the principal is unconstrained as to what information the agent can be induced to acquire and report honestly. When the agent is risk neutral, and a) is not asked to learn too much, b) can acquire information sufficiently cheaply, or c) can face sufficiently large penalties, the principal can attain the first-best outcome. We discuss the effect of risk aversion (on the part of the agent) and characterize the second-best contracts.
\end{abstract}
\keywords{Moral hazard, Information acquisition, Rational inattention, Information design}\\
\jel{D81; D82; D83; D86} 

\newpage

\section{Introduction}

People buy advice: investors pay for stock picks, politicians and executives in firms employ advisors, and bettors at the race track ask for winners. In some situations this advice can be backed up with hard, verifiable evidence; whereas in others advice is merely cheap talk and honesty is supported only by the advisor's incentives to be truthful. This paper studies the latter situation: we analyze a contracting problem in which a principal hires an agent to acquire unverifiable evidence, which cannot be credibly disclosed or contracted upon.\footnote{This is the key difference between this paper and \cite{rappoport2017incentivizing}, who explore a similar problem but specify that evidence is observable and contractible.}


In our model, it is costly for the agent to acquire information, and he has significant freedom in his learning: he may choose any distribution over posterior beliefs whose mean is the prior. Although the evidence an agent acquires is non-contractible, in our main analysis, we assume that the true state is. Under this assumption, we begin by observing that any contract induces a decision problem for an agent. This allows us to show that the principal can implement any feasible learning: she can write a contract such that the agent is willing to learn precisely as desired \textit{and} report honestly. That is, the agency problem does not impede the principal's ability to acquire information. This result stands in stark contrast to the classical setting (\cite{holmstrom1979moral}), wherein not all effort levels can be implemented in the second-best world.

Next, we show that the required incentives for the agent's learning produce a number of conditions whose structure allows us to simplify the principal's problem. For any state, each message contingent transfer in that state can be written as the difference between the transfer paid in that state for a ``benchmark message'' and a constant that depends only on exogenous values and the posteriors themselves. Not only do the relative incentives completely pin down the agent's optimal learning, but the converse is also true: the agent's optimal learning specifies the relative incentives. 


We solve for the cheapest contracts that induce the agent to acquire the desired information \emph{and} report his findings truthfully. As in the classical moral hazard environment, there is a natural benchmark in our model: the first-best problem in which learning (our analog of effort) is observable and contractible. We show that when the agent is risk neutral and negative transfers are allowed, any distribution over posteriors can be implemented at the first-best cost, even in our main setting with hidden learning and unverifiable evidence. Moreover, this holds even if the agent may exit the relationship after acquiring information, which renders the ``selling the project to the agent'' contract generically ineffective. This highlights another essential difference between the canonical setting and ours. In the classical setting, the possibility of an interim exit allows the agent to accrue rents. 

If negative transfers are forbidden (limited liability) and the outside option is sufficiently low, the principal cannot efficiently acquire information through the agent. Nevertheless, we show that optimal incentives take simple forms in a number of cases. We provide a full characterization of the optimal contract when the agent's outside option is sufficiently small. There, it is only the limited liability constraint that binds, 
which allows us to pin down the optimal contract for any desired distribution over posteriors. We also fully characterize optimal implementation with an arbitrary outside option and limited liability 
in the binary-state case when the agent is risk neutral. In particular, implementation is efficient if and only if the agent is not asked to learn too much (in relation to her cost of acquiring information and outside option).

We also show that the agent's risk aversion introduces inefficiencies: providing incentives for the agent to learn requires that he be exposed to risk, which is surplus destroying when the agent is risk averse. Similar to the classical setting, we establish that with only an \textit{ex ante} participation constraint, the agent gets zero rents. On the other hand, the possibility of an interim exit generically grants the agent surplus: there, the principal trades off conceding rents with better risk sharing.

We finish this section by reconciling our main setting's theoretical predictions with what we see in practice and also discuss related literature. After that, Section \ref{model} lays out the model before Section \ref{principal} states the principal's problem and discusses the first-best benchmark. Section \ref{preliminary} presents some preliminary results, and  Sections \ref{mr1} and \ref{mr2} contain the main results in the absence and presence of limited liability constraints, respectively. We wrap things up in Section \ref{discus}.

\subsection{Buying Opinions in Practice} \label{inpractice}

It seems indisputable that one essential assumption of our model--the softness and non-contractibility of the agent's findings--captures reality in some settings. Consider, for instance, talent scouting in sports. Although teams have hard evidence about prospects (goals scored, batting average, shooting percentage etc...), they nevertheless send scouts to obtain soft (unquantifiable) information: a report from an FC Barcelona scout describes a player's running style, balance, and control, among other things (\cite{barca}). The scout writes about the player's positioning, ``Excellent. It is undoubtedly his best quality. He is always where he should be...'' 

On the other hand, our assumption that the true state is contractible is more difficult to justify. Several comments are in order. First, if the true state is observable to the principal but not contractible, then as long as the principal has deep pockets, \cite{kleinwhit} show that the contracts we construct in this paper can be approximated by a mediated protocol, in which the agent occasionally acquires close to full information and the principal is penalized if caught in a (probable) lie.

Second, it may be that even the principal does not observe the realized state. If the principal has no other source of information about the state other than the agent then the situation is hopeless: the agent must be provided with differential incentives in order to learn and report honestly. However, it is reasonable in many cases that the principal does have access to information about the state other than that provided by the agent. For example, in professional baseball, once a prospect is identified by one of its scouts, a team will bring the player in to its training facility and evaluate him further directly.

The outcome of this sort of evaluation is subjective and private; however, the ultimate decision by the principal is not. Given this, one common contract is one that conditions the agent's reward on the principal's action. Baseball scouts receive bonuses if a prospect they recommended makes it to a team. Headhunters are rewarded if they identify a candidate that is hired--it is common for firms to pay recruiters a fraction of a hired worker's salary (crucially, \textit{if they are hired}, i.e., make it past the firm's final screening). In the supplemental appendix, we explore an example in which the principal obtains a private signal and the agent's payment is conditioned on the principal's action. There, we point out that i. not all distributions over posteriors may be implementable; ii. the agent may accrue greater rents; and iii. the principal's behavior in her decision problem may be distorted, a new variety of inefficiency.


\subsection{Related Literature}

Our study belongs to the literature on delegated expertise, pioneered by \cite{lambert1986executive}, \cite{demski1987delegated} and \cite{osband1989optimal}, in which a principal hires an agent to collect payoff relevant information. The central theme of this literature is incentive design for effective information acquisition and communication. 

There are five recent papers that are close to ours: the already referenced \cite{rappoport2017incentivizing}, who also study contracting for flexible information acquisition, but where the posterior generated by the agent’s choice of distribution is verifiable and contractible;\footnote{\cite{bizzotto2020information} consider a similar problem. However, they only allow the agent to deviate to a ``default'' distribution, instead of any Bayes-plausible distribution. Also related is \cite{yoder2022designing}, who generalizes the two-state (risk-neutral agent) environment of \cite{rappoport2017incentivizing} by incorporating a screening problem: the agent's marginal cost of acquiring information ($\kappa$) is his private information. \cite{wang2022contracting} subsequently generalizes this to \(n\) states.}
\cite{zermeno2011}; \cite{sharma}; \cite{clark2021contracts}; and \cite{muller}. \cite{zermeno2011} and \cite{clark2021contracts} both explore contracting environments in which both information acquisition and decision making are delegated to an agent. \citeauthor{zermeno2011}'s focus is the interaction between the variables on which the transfer schemes can depend and whether contracts specify transfer scheme menus. \cite{clark2021contracts} show that any Pareto-optimal contract can be decomposed into a fraction of output, a state-dependent transfer, and an optimal distortion.

\cite{sharma} and \cite{muller} are especially related. The former explores a two-state version of our environment with a risk-neutral agent and limited liability (and a low outside option). Our Proposition \ref{llbinds} is; therefore, a generalization of their elegant characterization result (Theorem 1). The latter work introduces a strikingly useful object--the ignorance equivalent--for studying rational inattention problems. As an application, they show that a principal can efficiently acquire information through a risk-neutral agent with only an \textit{ex ante} participation constraint.

\cite{carroll2019robust} studies a robust contracting problem in which the principal has limited knowledge about how the agent can learn and evaluates each possible contract by its worst-case guarantee. In \cite{haefner2022on} the agent acquires information to help the principal decide how much she should invest in a project. The distribution over posteriors and its cost are primitives of the model, and the agent's report of the realized posterior is unverifiable. Their focus is on finding the optimal contract--which can depend on the report and the outcome of the project--that motivates the agent to conduct the experiment and report truthfully.\footnote{\cite{terovitis2018motivating} tackles a similar problem. In his framework, the outcome is deterministically pinned down by the action and state, and the decision is delegated to the agent.} \cite{strategicfore} study the problem of a principal that designs a dynamic mechanism (without transfers) to identify a competent forecaster.

\cite{gromb2007collusion} consider a problem of delegated expertise with two agents, where the agents may collude among themselves or with the principal. In their model, the state space is binary, and the agent is restricted to a fixed message space containing two signals whose meaning is common knowledge. Their one-agent/one-signal case is similar to our model: the agent is risk neutral and protected by limited liability, the compensation can be conditioned on both the report and the realized state, and incentives must be provided for the agent to gather information and report truthfully. \cite{chade2016delegated} study a dynamic model of contracting for information acquisition in a two state-two (fixed) signals environment. The more effort the agent exerts, the more informative the signal he acquires. They assume that the realized signals are contractible, but the true state is not. 


Since in our model every contract induces a decision problem with a posterior separable cost of the agent, our work is naturally related to the rational inattention literature pioneered by \cite{sims1998stickiness, sims2003implications}. To analyze the agent's problem, we use insights from \cite{caplin2022rationally}. \cite{mackowiak2021rational} provides an excellent review of this literature that covers both theory and applications. 

Our principal also needs to elicit information from the agent, which connects our paper to the belief elicitation literature. Indeed, our transfer scheme is a scoring rule. The most important distinction is that the beliefs in our work are endogenously determined through the agent's learning. While the papers in that literature study what scoring rules induce truthful reporting (\cite{gneiting2007strictly} and \cite{schlag2015penny} are good surveys) and what properties of a state distribution can be elicited (see \cite{lambert2019elicitation} and references therein), our focus is on deriving incentive contracts that induce the agent to learn \emph{and} report truthfully.

Finally, because we study the motivation of an agent to acquire costly and unverifiable information, our work also connects to the moral hazard literature. In the canonical moral hazard problem (see, for example, \cite{mirrlees}, \cite{holmstrom1979moral}, \cite{grossman1983analysis}, and \cite{holmstrom1987aggregation}), the agent is impelled to exert costly effort that yields some (distribution over) output; whereas in ours, he must be coerced into choosing a much more complicated object (a particular probability distribution) then reporting honestly.

\section{The Model}\label{model}

There is an unknown state of the world $\theta \in \Theta$, where $\left|\Theta\right| = n < \infty$; and both principal and agent share a common full support prior $\mu \in \Delta\left(\Theta\right)$. 
The principal (she) has a continuous (reduced-form) payoff function over posteriors \(\mathbf{x} \in \Delta\left(\Theta\right)\), \(V\left(\mathbf{x}\right)\).\footnote{This could correspond, for example, to a decision problem faced by the principal.} The principal cannot acquire information herself but instead must rely on the assistance of an agent (he), who acquires information by conducting a costly experiment. As shown in \cite{kam}, this is equivalent to him choosing a distribution over posteriors \(F \in \Delta \Delta \left(\Theta\right)\) that is Bayes-plausible: \(\mathbb{E}_F[\mathbf{x}] = \mu\). The agent's cost of acquiring \(F\), denoted by \(C(F)\), is posterior separable \`{a} la \cite{caplin2022rationally}; that is,
\[C\left(F\right) = \kappa \int_{\Delta \left(\Theta\right)} c\left(\mathbf{x}\right) \, dF\left(\mathbf{x}\right) \text{ ,}\]
where $\kappa > 0$ is a scaling parameter, $c \colon \Delta \left(\Theta\right) \to \mathbb{R}_{+}$ is a strictly convex and twice continuously differentiable function bounded on the interior of $\Delta\left(\Theta\right)$, and $c\left(\mu\right) = 0$.\footnote{This class of information costs includes the entropy-based cost function (see e.g. \cite{sims1998stickiness, sims2003implications}, and \cite{matvejka2015rational}); the log-likelihood cost of \cite{costofinfo}; the neighborhood-based cost function studied by \cite{hebert2021neighborhood}; and the quadratic (posterior variance) cost function.}

After acquiring information, the agent sends a message to the principal. The true state is eventually observable to both parties after the principal’s value is realized and can be contracted upon. A contract specifies the set of messages available to the agent, and a transfer paid to the agent which can be contingent on both the realized state and the message sent. Formally, the principal proposes a pair $\left(M, t\right)$ consisting of a compact set of messages $M$ available to the agent, and a transfer $t \colon M \times \Theta \to \mathbb{R}$ ($t \colon M \times \Theta \to \mathbb{R}_+$ when the agent is protected by limited liability). 
We assume the principal's payoff is quasi-linear in the transfer. The agent's payoff is additive separable in his utility from the transfer and the cost of acquiring information, and he values the transfer according to a continuously differentiable, concave, and strictly increasing function $v$, with $v\left(0\right) = 0$. To ease presentation, transfer $t$ is expressed in utils. 

We further assume the agent has access to an outside option of value $v_0 \geq 0$, and that there are two chances for him to leave with his outside option: he can choose not to accept the contract, or walk away after acquiring information by reporting nothing. In doing so, the agent sends the ``null message,'' $\varnothing$.\footnote{More generally, the agent's outside option could derive from some salvage value for information that is an arbitrary upper semicontinuous function of the posterior $p\left(\mathbf{x}\right)$. In the supplementary appendix, we explain that this does not alter our analysis in a meaningful way.}

Unless otherwise noted, we assume throughout that the principal suffers a penalty that is strictly greater than \(v_{0}\) if the agent takes his outside option.\footnote{This captures the cost of hiring a new agent, for example.} This ensures that it is not optimal for the principal to offer the agent a contract in which he ever takes his outside option. In the discussion following Theorem \ref{dramatic}, we argue that our framework can accommodate ``shoot the messenger'' contracts--in which the agent is asked to exit the relationship with positive probability on path--with ease.

The timing of the game is as follows:
\begin{enumerate}[label={(\roman*)},noitemsep,topsep=0pt]
    \item The principal proposes a contract $\left(M,t\right)$;
    \item If the agent does not accept, the game ends, and the agent and principal receive \(v_0\) and \(V\left(\mu\right)\), respectively; otherwise the agent chooses a Bayes-plausible distribution \(F\), from which a posterior $\mathbf{x} \in \Delta \left(\Theta\right)$ is drawn and privately observed by the agent;
    \item The agent chooses whether to report. If he reports, he sends a message $m \in M$; and if he does not report, he takes his outside option \(v_{0}\) (and the principal observes the null message);\footnote{If the optimal contract is such that the null message, \(\varnothing\), is off-path (as it will be provided the principal incurs a sufficiently large cost from the agent exiting the relationship), the principal obtains \(V\left(\mu\right)\). If \(\varnothing\) is on-path, corresponding to posterior \(\mathbf{x}'\), the principal gets \(V\left(\mathbf{x}'\right)\).}
    \item Payoffs accrue: given belief \(\mathbf{x}\left(m\right)\) induced by message \(m\), the principal gets \(V\left(\mathbf{x}\left(m\right)\right) - \mathbb{E}_{\mathbf{x}\left(m\right)}v^{-1}\left(t\left(m,\theta\right)\right)\) and the agent $\mathbb{E}_{\mathbf{x}\left(m\right)}t\left(m,\theta\right) - c\left(F\right)$. 
\end{enumerate}

\section{The Principal's Problem}\label{principal}
\subsection{The First Best Benchmark}
Denote the set of Bayes-plausible distributions over posteriors by $\mathcal{F}(\mu)$. It is a convex and compact subset of $\Delta \Delta \left(\Theta\right)$. If the principal controlled the information acquisition, she would solve
\[\max_{F \in \mathcal{F}(\mu)} \int \left(V - \kappa c\right) \, dF \text{ ,}\]
which is a linear functional of \(F\), guaranteeing the existence of a maximizer. In our context, ``first best'' refers to the situation where the principal can observe the distribution over posteriors chosen by the agent, so the principal can specify transfer $t \colon \Delta \Delta \left(\Theta\right) \to \mathbb{R}_{+}$. When the distribution is observable, the following contract implements any distribution \(F\) and is optimal: the principal pays the agent precisely the amount that makes him indifferent between learning and walking away with his outside option if and only if the agent acquires \(F\). Otherwise, the principal pays the agent nothing. Evidently, the transfer is never strictly negative, and the agent is willing to acquire \(F\). Therefore, at the first best, the principal's cost of acquiring information is \(v^{-1}\left(C(F) + v_0\right)\).

\subsection{The Contracting Problem}
A contract must guarantee that the agent chooses the right distribution and reports honestly. 
Without loss of generality, every message contained in the message space \(M\)--except for the null message \(\varnothing\)--uniquely identifies a posterior in the support of \(F\),\footnote{The support of a distribution, denoted by \(\supp{(\cdot)}\), is the smallest closed set that has probability one.} and hence \(M = \supp(F) \cup \{\varnothing\}\). 
We say that a distribution \(F\) is \emph{implementable} if choosing \(F\) and reporting truthfully is an optimal strategy for the agent following some contract offer. 
Equivalently, the contract $(M,t)$ \emph{implements} \(F\). \(F\) can be \emph{implemented efficiently} if it can be implemented at the first-best cost.

Any contract \((M,t)\) offered to the agent produces a \emph{value function}
\[W(\mathbf{x}) \coloneqq \max_{m \in M}\mathbb{E}_{\mathbf{x}}\left[t\left(m,\theta\right)\right] - \kappa c\left(\mathbf{x}\right) \text{ ,}\]
which is the highest payoff the agent can obtain--if he accepts the contract and does not walk away after acquiring information--when his posterior is \(\mathbf{x}\). By construction, \(W\) is continuous and piecewise strictly concave. The agent chooses a distribution over posteriors to maximize his \emph{ex ante} value. 

\cite{caplin2022rationally} observe that the agent's optimal behavior corresponds to the hyperplane that is tangent to the concavified value function at the prior \(\mu\).\footnote{The concavified value function is the pointwise lowest concave function that majorizes the value function.} We denote this hyperplane by \(\mathcal{H}\) and sometimes refer to it as the \emph{concavifying hyperplane}. As is standard, we can identify this supporting hyperplane $\mathcal{H}$ with an affine function $f_\mathcal{H}\left(\mathbf{x}\right) \colon \Delta (\Theta) \to \mathbb{R}$. This hyperplane is the central object in our contracting problem under study: as we will shortly discover, the principal's implementation problem is essentially one of choosing this hyperplane, which pins down the required transfers. The agent's optimal \emph{ex ante} value is given by \(f_{\mathcal{H}}(\mu)\). 

\cite{caplin2022rationally} point out that the optimal posteriors are the points at which this hyperplane intersects \(W\); we denote the set of such points by \(P_{(M,t)}\):
\[P_{(M,t)} \coloneqq \left\{\mathbf{x} \in \Delta\left(\Theta\right)\colon f_\mathcal{H}\left(\mathbf{x}\right) = W\left(\mathbf{x}\right)\right\}\text{.}\]
By construction, at every $\mathbf{x} \in P_{(M,t)}$, it is optimal for the agent to report the realized posterior honestly. Therefore, a necessary condition for a distribution \(F\) to be implemented by a contract $(M,t)$ is that $\supp(F) = P_{(M,t)}$.\footnote{Technically, this is incorrect: the necessary condition is that \(\supp(F) \subseteq P_{(M,t)}\). The stated equality anticipates Proposition \ref{housekeeping}, in which we argue that WLOG \(F\) has affinely-independent support.} This need not be sufficient for implementation: the contract must also prevent the agent from walking away at any point in the interaction. In particular, no matter what the realized posterior is, the agent cannot deviate profitably by taking his outside option without making a report; this requires 
\[\label{icorig}\tag{$IR-v_0$} f_\mathcal{H}\left(\mathbf{x}\right) \ge v_0 - \kappa c(\mathbf{x}) \quad \text{for all} \quad \mathbf{x} \in \Delta(\Theta) \text{ .}\]
As this constraint imposes restrictions \emph{ex interim}, we often call it the interim individual rationality constraint (or participation constraint).\footnote{The relevance of this constraint is illustrated in an example in Section \ref{impeg}.} It is stronger than the \textit{ex ante} participation constraint $f_{\mathcal{H}}(\mu) \ge v_0$.




Thus, 
\begin{lemma} \label{fcontrimp}
    A contract $(M,t)$ implements distribution \(F\) if and only if
    \begin{enumerate}[label={(\roman*)},noitemsep,topsep=0pt]
        \item \label{fci1} \textbf{Incentive Compatibility}: $\supp(F) = P_{(M,t)}$; and 
        \item \label{fci2} \textbf{Individual Rationality}: Constraint \ref{icorig} holds; and
        \item \label{fci3} \textbf{(Ex Post) Limited Liability}: if there is limited liability, $t(m, \theta) \ge 0$ for all $\theta \in \Theta$ and $m \in M$.
    \end{enumerate}
\end{lemma}

The individual rationality constraint \eqref{icorig} is similar to the (interim) limited liability constraint in \cite{rappoport2017incentivizing} (p.11)--since \cite{rappoport2017incentivizing} allow for contracting upon the realized posterior, their limited liability is of \emph{ex interim} nature. Indeed, \eqref{icorig} requires that the value of the agent at any posterior in the support of the desired distribution (i.e., at the interim stage) cannot be too low. Here, this is a consequence of preventing the agent from learning according to a different distribution and not making a report following some posterior (rather than the direct imposition of \cite{rappoport2017incentivizing}). We stick to the term ``individual rationality'' because it also ensures that the agent accepts the contract \emph{ex ante} and prefers not to send the null message \emph{ex interim}. Limited liability \ref{fci3} in our work imposes restrictions \emph{ex post}. This concern is absent from \cite{rappoport2017incentivizing}, as contracts there are not state-contingent. To streamline exposition, we frequently drop ``ex post'' and refer to this constraint merely as ``limited liability.''

To solve the principal's contracting problem, we adopt a two-step approach: first, for every implementable distribution \(F\), we solve the principal's cost minimization problem:
\[\label{objective} \tag{$\ddagger$} \min_{(M,t)}\mathbb{E}_{F, \mathbf{x}}\left[v^{-1}\left(t\left(\mathbf{x}, \theta\right)\right)\right] \text{ ,}\]
subject to \ref{fci1}, \ref{fci2}, and \ref{fci3} in Lemma \ref{fcontrimp}; denote its value by $\Gamma(F)$. Second, the principal chooses an implementable distribution \(F\) to maximize her payoff under agency, $\int V(\mathbf{x}) \, dF(\mathbf{x}) - \Gamma(F)$. Like most papers studying moral hazard, we focus on the first step.

\begin{figure}
    \centering
    \includegraphics[width=0.57\linewidth]{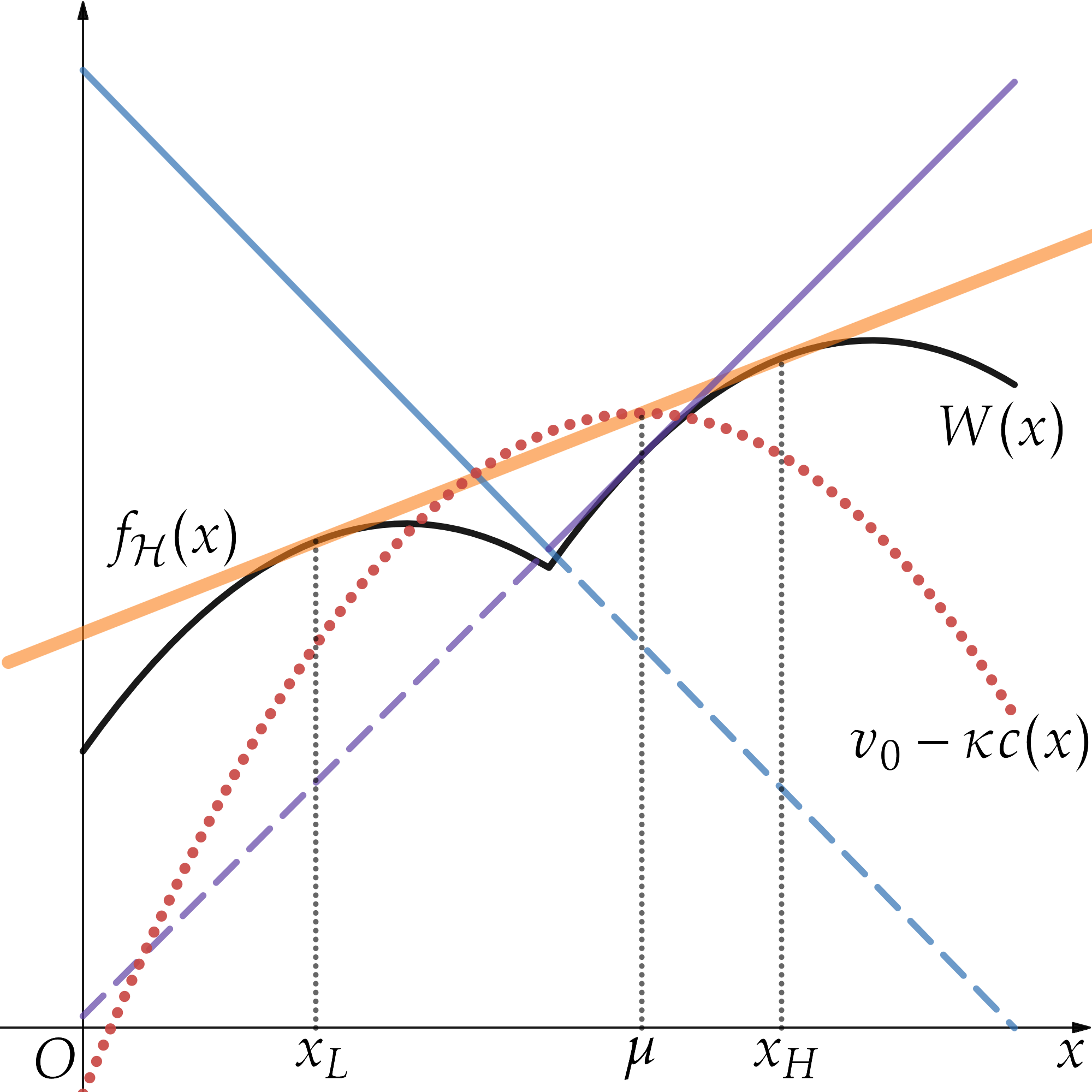} \\
    \vspace{40pt}
    \includegraphics[width=0.57\linewidth]{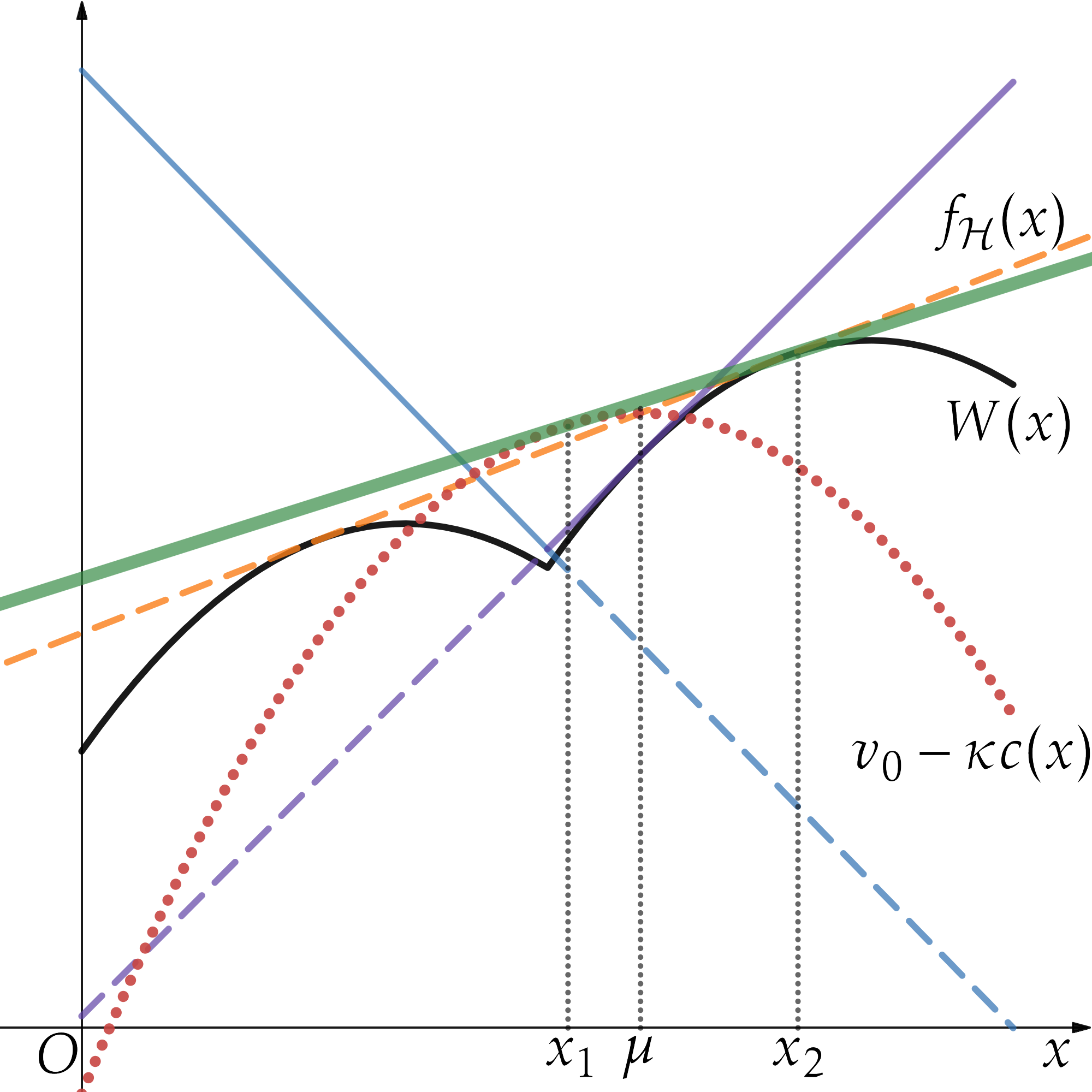}
    \caption{Contract \((\tilde{M},\tilde{t})\) fails to implement $\{x_L, x_H\}$, where $x_L = 1/4$ and $x_H = 3/4$, when $\mu = 3/5$, $\kappa = 2$, and $v_0 = 0.65$, and with quadratic cost: \(c(x) = (x-\mu)^2\). This contract satisfies the limited liability constraints, but interim IR \eqref{icorig} is violated.}
    \label{contract}
\end{figure}

\subsection{An Example} \label{impeg}

Let \(\Theta = \{\theta_L, \theta_H\}\) and \(x\) denote the posterior probability that the state is \(\theta_H\). Suppose the principal intends to implement distribution \(\{x_L, x_H\}\). Consider the contract \((\tilde{M},\tilde{t})\) where \(\tilde{M} = \{x_L, x_H, \varnothing\}\), \(x_L = 1/4\) and \(x_H = 3/4\), and $\tilde{t}$ is such that \(\tilde{t}(x_L, \theta_L) = 1.0125\), \(\tilde{t}(x_L, \theta_H) = 0\), \(\tilde{t}(x_H, \theta_L) = 0.0125\), and \(\tilde{t}(x_H, \theta_H) = 1\). The gross payoff to the agent from sending message \(m \in \{x_L, x_H\}\), as a function of posterior \(x\), is given by
\[\mathbb{E}_{x}[\tilde{t}(m, \theta)] = x \tilde{t}(m, \theta_H) + (1-x) \tilde{t}(m, \theta_L).\]
This example is illustrated in \autoref{contract}. The blue and purple lines depicts the gross payoff to the agent from sending \(x_L\) and \(x_H\), respectively. The maximum of these functions, net of the agent's learning cost, is the agent's induced value function, \(W\), depicted in black. The concavifying line \(f_{\mathcal{H}}\)--which pins down the agent's optimal learning--is in orange. Finally, the agent's net payoff from taking the outside option \(v_{0}\) is the dashed red curve.

If we ignore the individual rationality constraint, the contract \((\tilde{M},\tilde{t})\) induces the agent to acquire \(\{x_L, x_H\}\) and report truthfully because incentive compatibility holds: as shown in the upper panel of \autoref{contract}, the optimal posteriors for the agent are the points at which \(f_{\mathcal{H}}\) intersects \(W\), which are indeed \(\{1/4, 3/4\}\). However, the IR constraint is violated because \(f_{\mathcal{H}}\) does not lie entirely above \(v_0 - \kappa c\), which prevents the principal from implementing her desired distribution.\footnote{This example also illustrates that interim IR is stronger than \emph{ex ante} IR: as shown in the upper panel, \(f_{\mathcal{H}}(\mu) = v_0 - \kappa c(\mu) = v_0\), and hence \emph{ex ante} IR holds.} As shown in the lower panel of \autoref{contract}, the agent can profitably deviate by acquiring \(\{x_1, x_2\}\):\footnote{In this example, \(x_1 = 0.522\) and \(x_2 = 0.766\).} if \(x_1\) realizes he opts out by sending message \(\varnothing\) and takes his outside option, and he reports \(x_H\) if \(x_2\) realizes. This way, his \emph{ex ante} payoff is given by the line through \((x_1, v_0-\kappa c(x_1))\) and \((x_2, W(x_2))\) (depicted in green) evaluated at the prior \(\mu\), which is strictly higher than \(f_{\mathcal{H}}(\mu)\). 

\section{Preliminary Results}\label{preliminary}
We begin by arguing that any distribution over posteriors with support on \(n\) or fewer points can be implemented by some contract.

\begin{lemma} \label{finducible}
    If \(F\) is a distribution over posteriors with $|\supp(F)| \le n$ and $\supp(F) \subseteq \inter{\Delta(\Theta)}$, there exists a contract $(M,t)$ that implements \(F\), and the expected cost to the principal is finite.
\end{lemma}

The proof of Lemma \ref{finducible}, and all other proofs omitted from the main text, are collected in Appendix \ref{proofs}. For each \(F\) supported on \(n\) or fewer interior points of $\Delta(\Theta)$, because the cost function $c$ is bounded and differentiable on $\inter{\Delta (\Theta)}$, Lemma 2 of \cite{caplin2022rationally} guarantees that there is a decision problem such that \(F\) is optimal. Therefore, we can construct a contract with bounded transfers such that the agent finds it optimal to first acquire \(F\) then report the realized posterior truthfully. Moreover, by adding a finite constant to the transfer, we can make Constraint \ref{icorig} hold. Therefore, every such distribution can be implemented at finite cost.

Because the support of any extreme point of $\mathcal{F}(\mu)$ is on \(n\) or fewer affinely-independent points, any $F \in \mathcal{F}(\mu)$ can be obtained by randomizing over a set of contracts each of which implements a distribution with support on at most \(n\) affinely-independent points--consequently, any distribution whose support is on the interior of $\Delta \left(\Theta\right)$ can be induced at a finite expected cost. As it is cheaper for the principal to randomize first rather than implement \(F\) directly, it is without loss of generality for the principal to implement a distribution over posteriors with support on at most \(n\) affinely-independent points. 
\begin{proposition} \label{housekeeping}
    \begin{enumerate}[label={(\roman*)},noitemsep,topsep=0pt]
         \item \label{finitecost} Every $F \in \mathcal{F}(\mu)$ with $\supp(F) \subseteq \inter{\Delta(\Theta)}$ can be implemented at a finite cost.
        \item \label{wlog} Without loss of generality, the principal only implements distributions with support on at most \(n\) affinely-independent points.
    \end{enumerate}
\end{proposition}
By Proposition \ref{housekeeping} \ref{wlog}, we can restrict our attention to distributions over posteriors with support on $\{\mathbf{x}_1, \mathbf{x}_2, \dots, \mathbf{x}_s\}$, where \(n\) is the number of states, and $s  \leq n$. In our next result, we discover that incentive compatibility allows us to reduce transfers to a single variable for each state.

For each state $k = 1, \dots, n$, define $\Omega^k\left(i,j\right) \coloneqq t_i^k - t_j^k$ ($i, j = 1, \dots, s$), where \(t_i^k := t(\mathbf{x}_i, \theta_k)\) is the promised payment to the agent from sending message \(i\) in state \(k\). Accordingly, each $\Omega^k\left(i, j\right)$ specifies the difference between the payoff to the agent from sending any message $i$ versus message $j$ in state $k$. Importantly, because (on path) each message corresponds to a different posterior, the collection of differences $\left(\Omega^k\left(i,j\right)\right)_{k=1}^{n}$ captures the relative benefit to the agent from obtaining posterior $j$ rather than posterior $i$. 


\begin{theorem}\label{dramatic}
The relative incentives $\left(\Omega^{k}\left(i,j\right)\right)_{i,j = 1, \dots,s; \, k = 1, \dots, n}$ are completely pinned down by incentive compatibility.

Consequently, the principal's problem of optimally inducing a distribution over posteriors reduces to an \(n\)-variable optimization problem, where \(n\) is the number of states. For each state $k$, the principal fixes a benchmark message $j\left(k\right)$, then chooses $\left(t_{j\left(k\right)}^{k}\right)_{k = 1}^{n}$; the payoff to the agent from sending message $j\left(k\right)$ in state $k$.
\end{theorem}

Theorem \ref{dramatic} is reminiscent of the standard result that truthtelling only identifies relative payments in adverse selection settings. Here; however, the relative incentives are pinned down jointly by the optimality of the desired distribution \emph{ex ante} and truthful reporting \emph{ex interim}. As part \ref{fci1} of Lemma \ref{fcontrimp} states, incentive compatibility for the agent requires that the value function of the agent, $W$, intersects the concavifying hyperplane at the support points of the distribution over posteriors the contract aims to implement. Such a hyperplane pins down the transfers in each state for each posterior in the support of the agent's learning. Consequently, the principal's problem is equivalent to one of choosing a hyperplane, which is an \(n\)-variable optimization problem.\footnote{Framed in this manner, this theorem is closely related to Lemma 2 in \cite{caplin2022rationally}, which states that when constructing a decision problem the tangent hyperplane is arbitrary.}

That was a technical explanation, here is an economic one. Fix a desired distribution; if the agent wants to deviate by slightly increasing the probability of a message realization 
in a certain state, basic probability implies that there must be a commensurate decrease in the probability of another message realization in that state. At the optimum, no such local deviation in the agent's information acquisition strategy can be profitable. Hence for any two posteriors in the support of the desired distribution, any deviation of the sort described above must generate a marginal value to the agent equal to the marginal cost. Because the marginal value of varying the probability of a message realization in a state is determined by the transfer for sending that message in that state, this ``zero net marginal gain'' observation generates an equality that connects the transfers for sending two distinct posteriors in the support of the desired distribution in the same state.

Recall that we specified early on that the principal suffers a disutility greater than \(v_{0}\) should the agent take his outside option. This ensures that the principal does not want to replace one of the messages with the null message, i.e., have the agent exit the relationship, sending the null message with positive probability. In principle, if the principal is not hurt (severely) by the agent's exit, it could be optimal for the principal to write a contract in which the null message is sent with positive probability (inducing the desired posterior) thereby allowing the principal to save on paying the agent. By Theorem \ref{dramatic} the belief to which the null message corresponds pins down the other transfers. Thus, if the principal's penalty from an agent's exit is less than \(v_{0}\), one must check at most $s$ additional contracts (other than those in which the null message is never sent), in which the null message is sent after each belief, in turn.\footnote{In the supplementary appendix we discuss an example in which it is optimal for the agent to exit the relationship with positive probability.}

\section{Main Results I. No (\emph{Ex Post}) Limited Liability}\label{mr1}

\subsection{Risk-Neutral Agent} \label{nllrna}
We first assume that the agent is risk neutral; without loss of generality, \(v\left(\cdot\right) = \cdot\). Theorem \ref{dramatic} implies that choosing a contract is tantamount to choosing a concavifying hyperplane \(\mathcal{H}\). Recalling that \(f_{\mathcal{H}}\) is the function that identifies \(\mathcal{H}\), and the agent's value from acquiring information for the principal (from an \textit{ex ante} perspective) is \(f_{\mathcal{H}}(\mu)\). To implement distribution \(F\) efficiently, the principal must only pay the first best cost, namely \(v^{-1}(C(F) + v_0) = C(F) + v_0\). Hence, efficient implementation requires \(f_{\mathcal{H}}(\mu) = v_0\). Furthermore, for \eqref{icorig} to hold, the graph of \(f_{\mathcal{H}}\) must lie entirely above the graph of \(v_0 - \kappa c\). Then, because \(f_{\mathcal{H}}\) is affine and \(c\) is strictly convex,

\begin{observation} \label{rntangency}
    When the agent is risk neutral, a principal can implement a distribution efficiently if and only if \(f_{\mathcal{H}}\) is tangent to \(v_0 - \kappa c\) at \(\mu\).
\end{observation}

Applying Observation \ref{rntangency}, efficient implementation is equivalent to the following \(n\) conditions:
\[\label{icreduced} \tag{$IR-R$} t_j^k - t_j^n - \kappa c_k\left(\mathbf{x}_j\right) = - \kappa c_k\left(\mu\right) \quad \text{for all} \quad k = 1, \dots, n-1 \text{,}\]
and $f_{\mathcal{H}}(\mu) = v_0$. We are able to pick an arbitrary support point \(\mathbf{x}_j\) of \(F\), as implementability requires that at each \(\mathbf{x} \in \supp(F)\), the agent's value function \(W\) induced by the contract intersects the same supporting hyperplane (\(\mathcal{H}\)). 

\begin{figure}
    \centering
    \includegraphics[scale=.25]{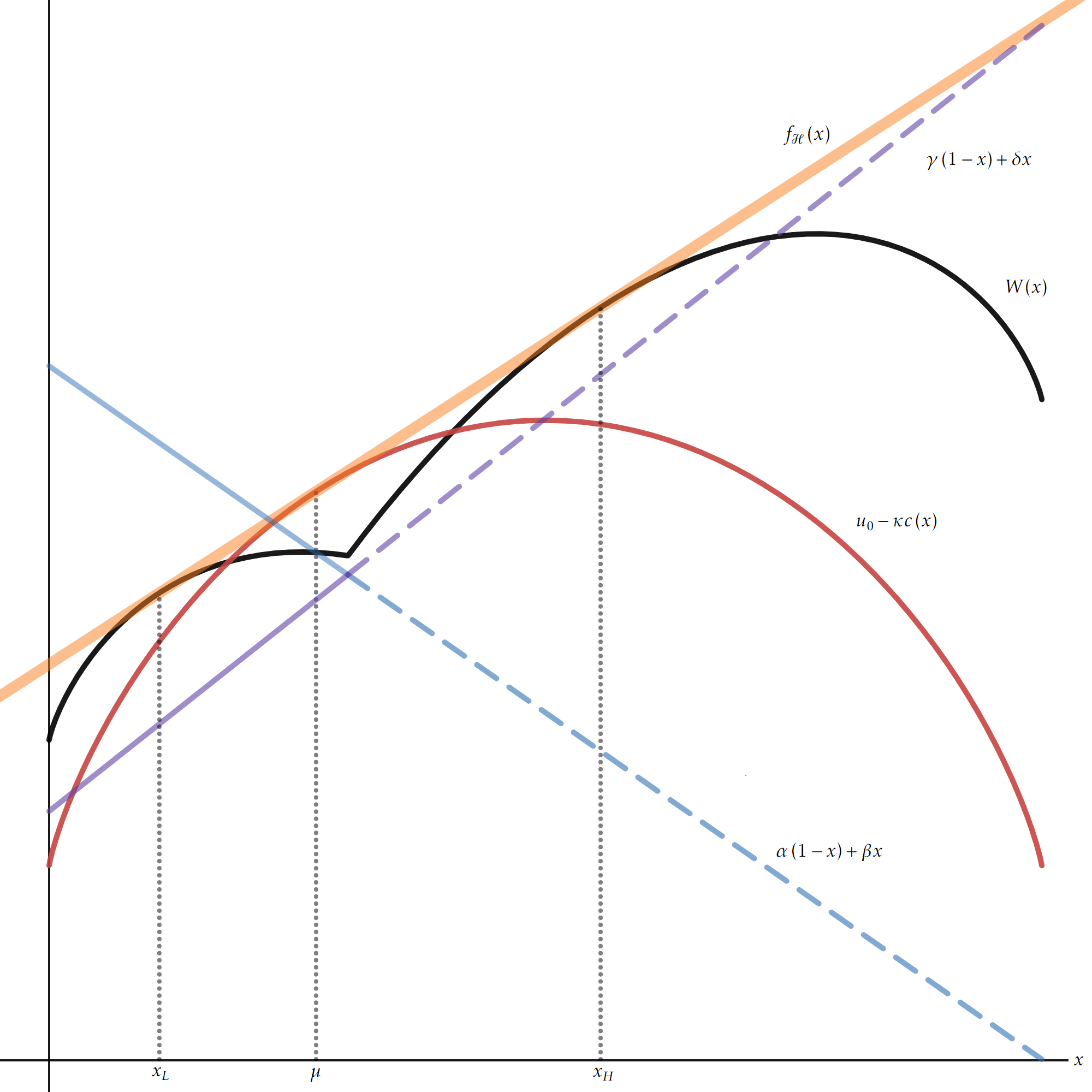}
    \caption{Efficient implementation of $\{x_L, x_H\}$, where $x_L = 1/9$ and $x_H = 5/9$, when $\mu = 1/\left(1+e\right)$, $\kappa = 1$, $v_0 = \log{\left\{9/(1+e)\right\}}$, and with entropy cost. This contract satisfies the limited liability constraints--as stated in Proposition \ref{effimp}, the specified ratio $v_0/\kappa$ is the minimum such ratio such that efficient implementation is feasible under limited liability.}
    \label{figeff}
\end{figure}

By Theorem \ref{dramatic}, the solution to this system, $\left(t^k_j\right)_{k=1}^{n}$, if it exists, identifies a contract. Accordingly, whether a distribution can be implemented efficiently boils down to whether the system of equations defined by the $n-1$ equations in Constraint \ref{icreduced} and $f_{\mathcal{H}}(\mu) = v_0$ has a solution. A solution always exists: 

\begin{proposition} \label{effinoll}
    If the agent is risk neutral and not protected by limited liability, every (feasible) distribution \(F\) with $\supp(F) \subseteq \inter \Delta (\Theta)$ can be implemented efficiently.\footnote{The supplementary appendix reveals that this holds even when there are uncountably many states.}
\end{proposition}
When there is no limited liability, the amount of incentive constraints is ``just right'' such that there exists a transfer scheme that delivers the right incentives and keeps the agent's surplus at his outside option. \autoref{figeff} illustrates this construction: the principal can always find a contract such that the concavifying hyperplane $f_{\mathcal{H}}$ (depicted in orange) of the agent's value function $W$ (in black) is a supporting hyperplane of the graph of $v_0 - \kappa c$ (the red curve), which is the agent's payoff from exiting the relationship. Therefore, the interim IR constraint \eqref{icorig} is always satisfied. 
The following \href{https://www.desmos.com/calculator/x0yorxyavo}{Interactive Link} illustrates the optimal contract (for an arbitrary binary distribution with support \(\left\{l,h\right\}\)) when there are two states, the agent's information acquisition cost is entropy-reduction, and his outside option is \(0\). 

It is instructive to compare Proposition \ref{effinoll} to Proposition 2 in \cite{rappoport2017incentivizing}, which states that when the realized posteriors are contractible (but the true state is not), efficient implementation is possible when the agent is risk neutral, even if he is protected by limited liability. As we discussed after Lemma \ref{fcontrimp}, their limited liability is of \emph{ex interim} nature, and is similar to our individual rationality constraint \eqref{icorig}. Consequently, one interesting way to interpret Proposition \ref{effinoll} is that so long as the realized state can be contracted upon, \cite{rappoport2017incentivizing}'s Proposition 2 still holds when the principal must elicit the agent’s belief instead of being able to contract on it. 

However, if we require \emph{ex post} limited liability, for some distributions and outside option values, efficient implementation cannot be achieved (irrespective of the interim IR constraint's presence).
In our model, to induce the agent to gather information, the transfers must be ``rewarding'' when the agent ``gets the state right'' and ``punishing'' when he is wrong. This gap between the two outcomes must be large enough to justify the cost of learning. Therefore, when \(v_{0}\) is small enough, to achieve an expected transfer of $\Gamma(F) = C(F) + v_0$, some ``punishing'' transfer(s) must be negative.

\subsubsection{Comparison to Classical Moral Hazard with an Interim Participation Constraint} \label{smhwiir}

A natural question is whether an analog of Proposition \ref{effinoll} holds if we add in an interim participation constraint to the canonical moral hazard problem. That is, is the ability of the principal to accommodate the interim participation constraint a special feature of our information acquisition problem or does it also hold in the classical environment?

As we show in the supplementary appendix, in the classical environment, the ability of the agent to exit the relationship after (privately) observing his output realization is tantamount to limited liability: clearly a contract cannot promise the agent a payoff less than his outside option for any (divulged) output. Moreover, because a contract must be incentive compatible, unless the principal implements the lowest possible effort, i.e., pays a constant wage, the agent must get strictly positive rents. 

\subsubsection{Selling the Project to the Agent?}

One might also wonder whether Proposition \ref{effinoll} is really needed. In the standard moral hazard problem, when the agent is risk neutral and there are no limited liability constraints, the principal can attain the first best by ``selling the project to the agent'' (henceforth the \textbf{STP} contract). In our setting,\footnote{We are also assuming here that the set of actions in the principal's decision problem is finite. The construction when the principal has infinitely many actions is analogous but more ungainly, so we omit it.} that corresponds to the principal writing the contract such that the agent's net utility as a function of his posterior $\mathbf{x}$ is $V\left(\mathbf{x}\right) - \kappa c\left(\mathbf{x}\right) - \tau$, where $\tau = f_{\mathcal{H}}\left(\mu\right) - v_{0}$. That is, the principal writes a contract so that the agent's problem, gross of the cost, is precisely that faced by the principal, then lowers the transfers to the agent uniformly to leave his \textit{ex ante} expected payoff equal to his outside option.

Already, the similarity between our interim IR constraint and (interim) limited liability suggests that this may not be possible (generically). Indeed, that is so in the standard moral hazard setting. This hunch is correct: when the agent can exit \textit{ex interim}, the principal cannot implement a distribution over posteriors at the first-best cost generically by selling the project. Recall that the optimal contract must be robust to double deviations in which the agent learns differently then takes her outside option with positive probability: \(f_{\mathcal{H}}\) must lie everywhere above \(v_0 - \kappa c\left(\mathbf{x}\right)\). Moreover, efficient implementation is even more demanding: \(f_{\mathcal{H}}\) must be tangent to $v_0 - \kappa c\left(\mathbf{x}\right)$ at $\mu$. As we show in the supplementary appendix, 
this tangency property is a non-generic property of the principal's information acquisition problem. 


\subsection{Risk-Averse Agent}
When the agent is risk averse, but unprotected by limited liability, characterizing the optimal contract is more involved. Fix an arbitrary benchmark message for all states, say $j$; the principal's payoff is strictly decreasing in each of the \(n\) control variables $\left(t_j^k\right)_{k = 1}^{n}$ and so the principal wants to set each one as low as possible. Unencumbered by limited liability, the lone constraint is \ref{icorig}, which necessarily binds (since otherwise, the principal could reduce the control variables). Thus,
\begin{observation} \label{ratangency}
    When the agent is risk averse, there exists an $\mathbf{x}^{*} \in \Delta \left(\Theta\right)$ such that $f_{\mathcal{H}}\left(\mathbf{x}\right)$ is tangent to $v_0 - \kappa c\left(\mathbf{x}\right)$ at $\mathbf{x}^{*}$.
\end{observation} 
Given this, solving for the optimal implementation of a distribution over posteriors \(F\) can be turned into an $n-1$ variable optimization problem by using the tangency conditions to substitute in for each $t_j^k$. This yields the principal an objective that is a function of $\mathbf{x}^{*}$.\footnote{More precisely, for each $\mathbf{x}^* \in \Delta\left(\Theta\right)$, $\left(t_j^k\right)_{k = 1}^{n}$ can be solved from the following $n$ equations: $t_j^k - t_j^n - \kappa c_k\left(\mathbf{x}_j\right) = - \kappa c_k\left(\mathbf{x}^{*}\right)$ for all $k = 1, \dots, n-1$, and $f_\mathcal{H}\left(\mathbf{x}^*\right) = v_0 - \kappa c\left(\mathbf{x}^*\right)$. Moreover, the relative incentives identified in Theorem \ref{dramatic} allow us to obtain the other transfers. Plugging the transfers into \eqref{objective}, the principal's objective can then be written as a function of $\mathbf{x}^*$.} Unless $\mathbf{x}^{*} = \mu$, which does not hold in general, the agent obtains positive rents. This finding is a consequence of the interim participation constraint, which requires that $f_{\mathcal{H}}$ lie above $v_0 - \kappa c$ everywhere. Otherwise--with only \textit{ex ante} IR--the agent would not obtain rents. Indeed, without the interim IR constraint, the lone constraint is the \textit{ex ante} participation constraint, which obviously binds; hence, $f_{\mathcal{H}}\left(\mu\right) = v_0$. Importantly, the agent's risk aversion engenders inefficiencies: the agent must be exposed to risk in order to acquire information and report honestly, which destroys surplus due to his risk aversion.

\begin{proposition} \label{riskaverse}
    Suppose the agent is risk averse and not protected by limited liability.
    \begin{enumerate}[label={(\roman*)},noitemsep,topsep=0pt]
        \item For every distribution over posteriors $F$ with support in \(\inter \Delta (\Theta)\), an optimal contract exists, and the transfers can be found by choosing $\mathbf{x}^* \in \Delta\left(\Theta\right)$.
         \item If the agent can exit \textit{ex interim}, he gets strictly positive rents unless $\mathbf{x}^* = \mu$. If the agent cannot exit \textit{ex interim}, he gets zero rents.
         \item Only the degenerate distribution of posteriors can be implemented efficiently.
    \end{enumerate}
\end{proposition}

In choosing $\mathbf{x}^*$, the principal optimally trades off between risk sharing and conceding rents: when a contract that makes the agent break even entails too much risk, moving $\mathbf{x}^{*}$ away from $\mu$ mitigates this risk. Then, although the agent receives strictly positive rents, implementing the new contract can be cheaper to the principal. This is reminiscent of the trade-off studied in Proposition 5 in \cite{rappoport2017incentivizing} though the exact mechanisms are different:\footnote{The resemblance stems from the fact that our interim IR constraint is similar to the (interim) limited liability constraint in \cite{rappoport2017incentivizing}, which leads to similar tangency conditions.} in their work, the most cost-efficient way for compelling the agent to choose the right distribution is to have the hyperplane determined by the wage contract (which, in their setting, is a function of the \textit{verifiable} posterior) to be tangent to the agent's value function. In our problem, averting double deviations to the outside option is what begets the tangency condition mentioned in Observation \ref{ratangency}. 

\section{Main Results II. (\emph{Ex Post}) Limited Liability}\label{mr2}

Throughout this section, we assume that the agent is protected by limited liability. In Subsection \ref{lowoo}, we solve for the optimal incentives when the agent's value for his outside option is sufficiently small. In Subsection \ref{rnagent}, we allow for an arbitrary outside option but impose that the agent is risk neutral.

\subsection{Low Outside Option}\label{lowoo}
For simplicity, we set $v_0 = 0$; it is not hard to see that all the results in this subsection go through for all sufficiently small $v_0 > 0$. By Theorem \ref{dramatic}, for any desired distribution, the relative incentives are identified. Consequently, for each state we can pinpoint a benchmark message that determines the lowest payment.

\begin{lemma} \label{lowest}
    For every state $k = 1, \dots, n$, there exists $j^*(k)$ such that $t\left(\mathbf{x}_{j^*(k)}, \theta_k\right) \le t\left(\mathbf{x}_{i}, \theta_k\right)$ for all $i = 1, \dots, s$.
\end{lemma}


Lemma \ref{lowest} allows us to completely identify the optimal transfers when the agent's outside option is sufficiently low.

\begin{proposition} \label{llbinds}
Suppose $v_0 = 0$, and the agent is protected by limited liability. Then for each state $k = 1, \dots, n$, there exists $j^*(k)$ such that $t\left(\mathbf{x}_{j^*(k)}, \theta_k\right) = 0$, and all other transfers are nonnegative and determined by the relative incentives identified in Theorem \ref{dramatic}. 
\end{proposition}

Proposition \ref{llbinds} is intuitive: for a sufficiently small outside option, Constraint \ref{icorig} always holds, so the transfer scheme is pinned down by optimal learning and limited liability. Optimal learning leaves, for each state, one degree of freedom to the principal; and to satisfy limited liability, the best that the principal can do is to find the smallest transfer in each state and set it to zero. The following \href{https://www.desmos.com/calculator/ndcajdkb3c}{Interactive Link} illustrates the optimal contract (for an arbitrary binary distribution with support \(\left\{l,h\right\}\)) when there are two states, the risk-neutral agent's information acquisition cost is entropy-reduction, and his outside option is \(0\).


\subsection{Risk-Neutral Agent}\label{rnagent}

Now, we dispense with the assumption that the outside option is small--\(v_{0}\) can take any value. For expository ease, we start with the two state case and then argue that our results generalize when there are more than two states.

\subsubsection{Two States}
When there are just two states, $\Theta = \left\{\theta_1,\theta_2\right\}$. By Proposition \ref{housekeeping} \ref{wlog}, we can identify a distribution by its support $\{x_L, x_H\}$. Our first result characterizes the distributions over posteriors that a principal can implement efficiently; \textit{viz.}, at the first-best cost. Defining
\[\eta\left(x_L,x_H\right) \coloneqq \max\left\{-\mu c'\left(\mu\right) - c\left(x_H\right) + c'\left(x_H\right) x_H, \left(1-\mu\right)c'\left(\mu\right) - c\left(x_L\right)- \left(1-x_L\right) c'\left(x_L\right)\right\} \text{ ,}\] we have
\begin{proposition}\label{effimp}
The principal can implement $\left\{x_L, x_H\right\}$ efficiently if and only if $v_0/\kappa \geq \eta\left(x_L,x_H\right)$.
\end{proposition}
Proposition \ref{effimp} states that a given distribution can be implemented efficiently if and only if either the agent has a sufficiently high outside option, or he can acquire information sufficiently cheaply. Intuitively, efficient implementation under limited liability requires that (1) the average payment to the agent net of the cost of information acquisition is $v_0$ (and Proposition \ref{effinoll} shows that \eqref{icorig} can be satisfied by ``tilting the hyperplane''), and (2) the payments from sending the ``wrong message'' in both states cannot be negative. (1) and (2) together imply that the differential payments between the ``right message'' and the wrong one cannot be too large compared to $v_0$. The differential payments are exactly the relative incentives identified in Theorem \ref{dramatic}, which are pinned down jointly by the desired distribution and model primitives (i.e., \(\kappa\) and \(c\)). This produces the inequality in Proposition \ref{effimp}.

Consequently, as $v_0$ gets larger, efficient implementation is easier. Furthermore, for a smaller $\kappa$ or a (Blackwell) less informative distribution, the principal only needs smaller differential payments to incentivize information acquisition, which also makes efficient implementation easier to achieve. Therefore, the left-hand side of Proposition \ref{effimp}'s necessary and sufficient condition is strictly decreasing in the information cost parameter $\kappa$. Moreover, the function $\eta$ is decreasing in $x_L$ and increasing in $x_H$. This suggests the following corollary:
\begin{corollary} \label{coreff}
\begin{enumerate}[label={(\roman*)},noitemsep,topsep=0pt]
    \item \label{corefi} For any pair of posteriors $\left\{x_L, x_H\right\}$ with $0 < x_L \leq \mu \leq x_H < 1$, if $v_0/\kappa$ is sufficiently large, $\left\{x_L, x_H\right\}$ can be implemented efficiently.\footnote{If $c'\left(0\right)$ and $c'\left(1\right)$ are finite, this is true for any $0 \leq x_L \leq \mu \leq x_H \leq 1$.} 
    \item \label{blackwellmon} Efficient implementation is \textit{monotone} with respect to the Blackwell order: if $\left\{x_L, x_H\right\}$ can be implemented efficiently, then any distribution that corresponds to a less informative experiment can be implemented efficiently.
    \item \label{blackwellunin} If $v_0 > 0$ then any distribution that corresponds to a sufficiently uninformative experiment can be implemented efficiently.
\end{enumerate}
\end{corollary}
In the canonical moral hazard problem with a risk-averse agent, no matter what outside option the agent has, only the lowest action can be implemented efficiently. Corollary \ref{coreff} has a flavor of that classical result: efficient implementation is possible whenever the agent is not asked to learn too much. For distributions more spread out than some set of threshold distributions; however, positive rents must be provided to the agent. To implement such distributions efficiently it must be that the relative incentives are high enough for the agent to acquire that much information and so when \(v_{0}\) is small limited liability is always violated.

When the first-best implementation of $\left\{x_L,x_H\right\}$ is infeasible, there are three other possibilities, listed in our next proposition. Denoting $\gamma \coloneqq t_{2}^{1}$ the transfer from sending message $x_L$ in state $\theta_2$, and $\beta \coloneqq t_1^2$ the transfer from sending message $x_H$ in state $\theta_1$, we have
\begin{proposition}\label{rncharacterization}
One of the following must occur at the optimum. Either
\begin{enumerate}[label={(\roman*)},noitemsep,topsep=0pt]
    \item \label{rnci} $\left\{x_L,x_H\right\}$ can be implemented efficiently (and Constraint \ref{icorig} binds); or 
    \item $\left\{x_L,x_H\right\}$ cannot be implemented efficiently; and either
    \begin{enumerate}[noitemsep,topsep=0pt]
    \item \label{rncii} Constraint \ref{icorig} binds and $\beta = 0$; or
    \item \label{rnciii} Constraint \ref{icorig} binds and $\gamma = 0$; or
    \item \label{rnciv} Constraint \ref{icorig} does not bind and $\gamma = \beta = 0$.
\end{enumerate}
\end{enumerate}
\end{proposition}

\begin{figure}
\centering
\begin{subfigure}{.5\textwidth}
  \centering
  \includegraphics[scale=.14]{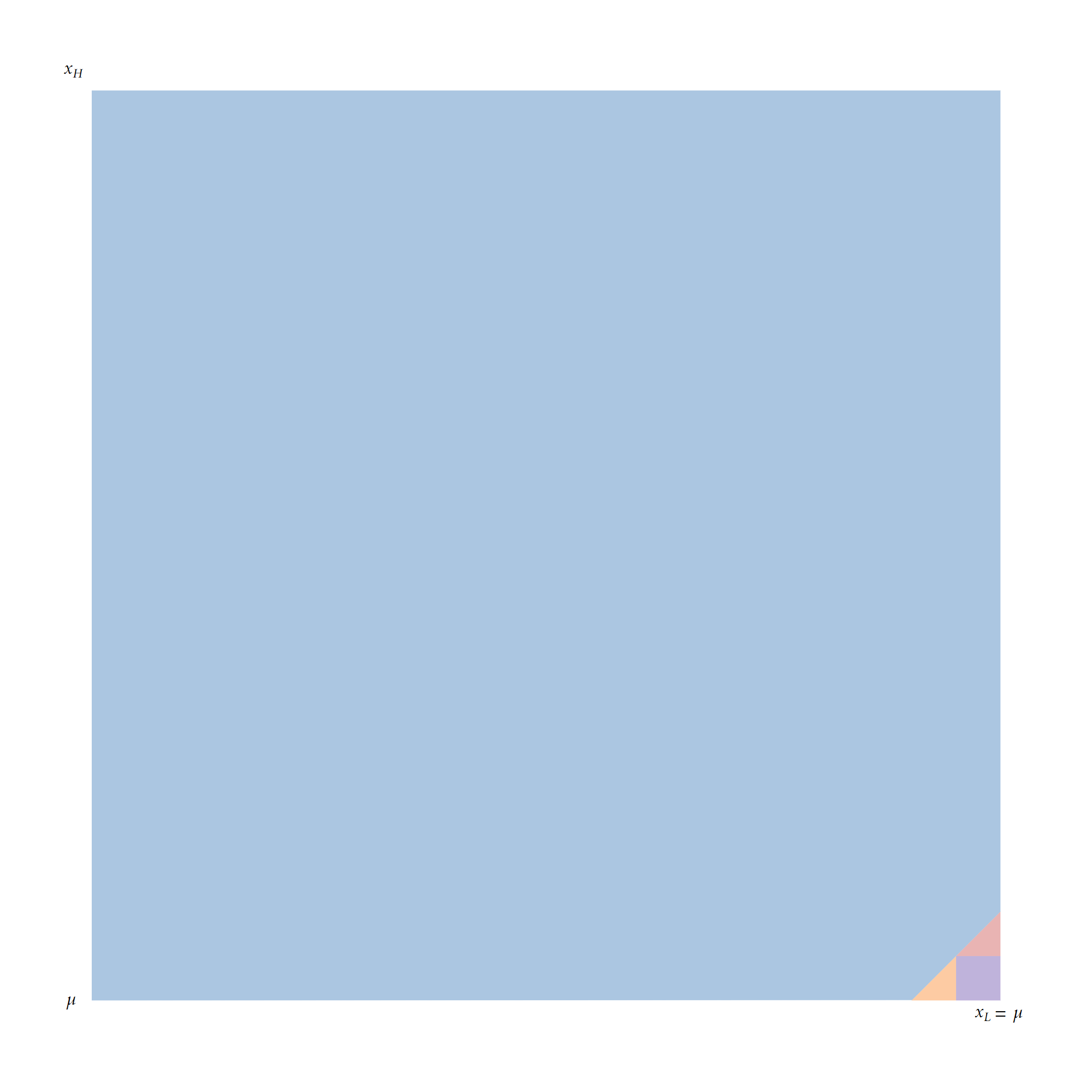}
  \caption{$v_0/\kappa = .05$.}
  \label{figsub1}
\end{subfigure}%
\begin{subfigure}{.5\textwidth}
  \centering
  \includegraphics[scale=.14]{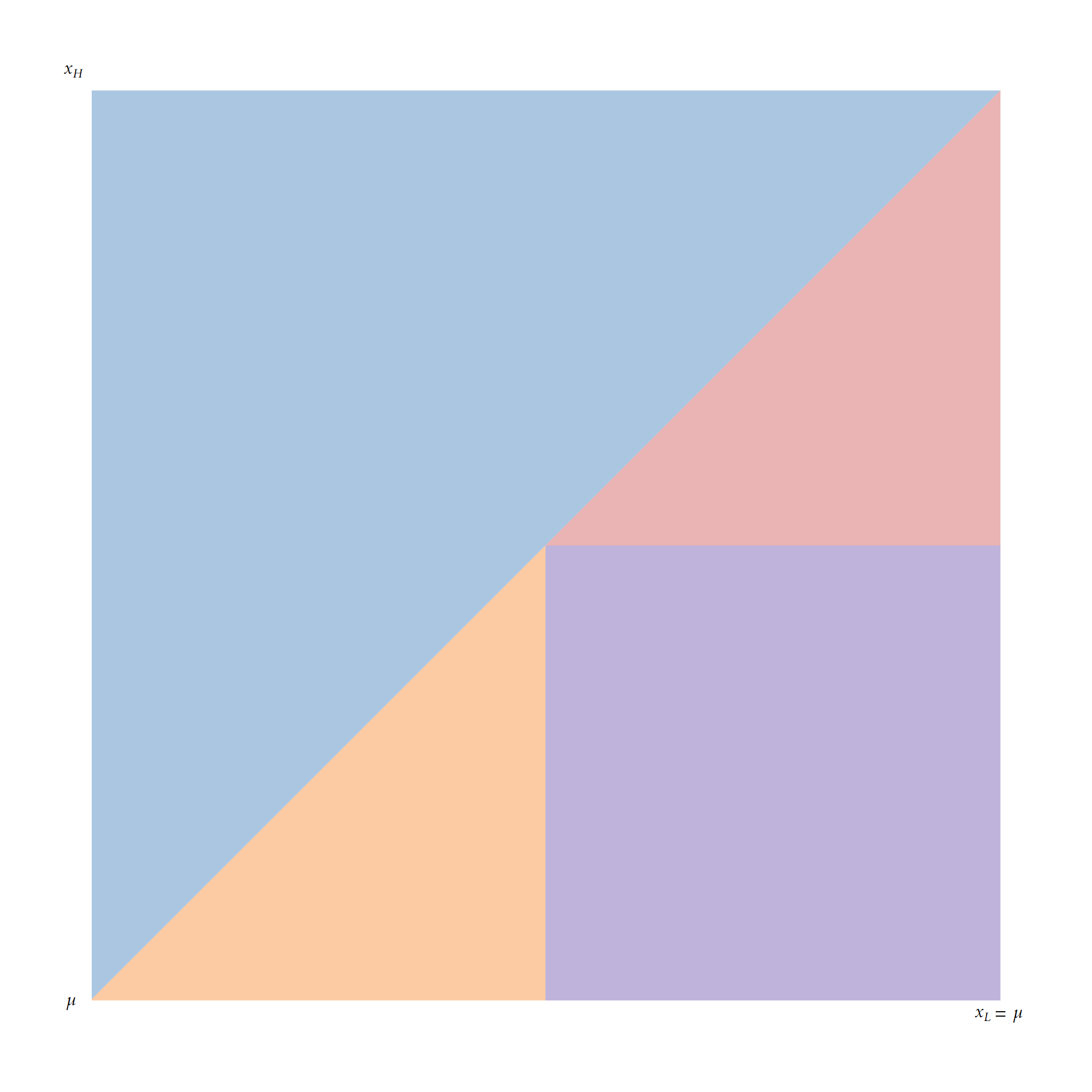}
  \caption{$v_0/\kappa = \log{\left(1/\left(1-\mu\right)\right)}$.}
  \label{figsub2}
\end{subfigure}
\par
\bigskip
\par
\begin{subfigure}{.5\textwidth}
  \centering
  \includegraphics[scale=.14]{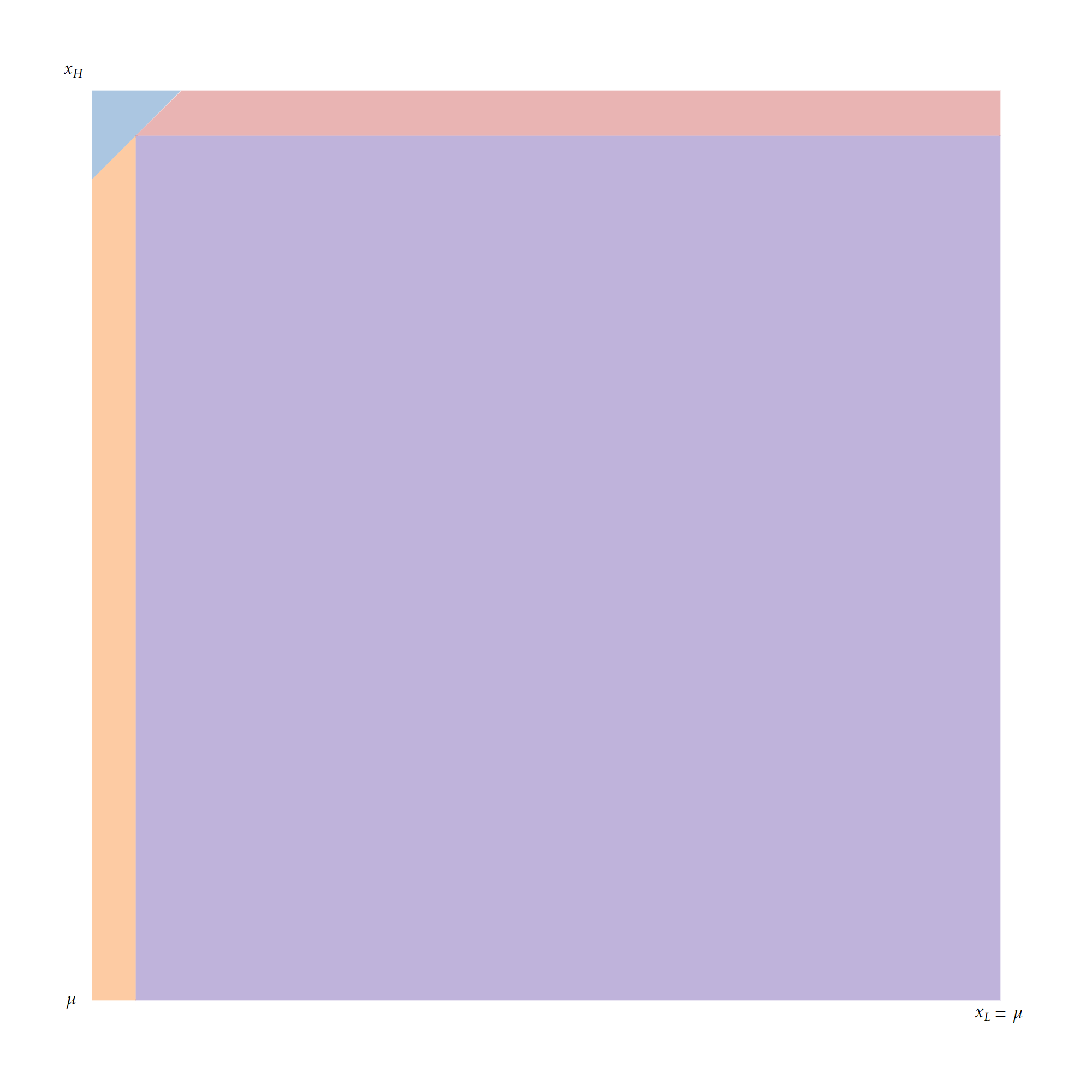}
  \caption{$v_0/\kappa = 3$.}
  \label{figsub3}
\end{subfigure}
\caption{\textbf{Implementation Regions for $\mu = 1/2$:} \(x_L\) is on the horizontal axis, ranging from 0 to \(\mu = 1/2\); and \(x_H\) is on the horizontal axis, ranging from \(\mu = 1/2\) to 1. Pairs $\left(x_L,x_H\right)$ in the purple region can be implemented efficiently, $\left(x_L,x_H\right)$ in the blue region are optimally implemented by $\gamma = \beta = 0$, $\left(x_L,x_H\right)$ in the orange region are optimally implemented by $\beta = 0$ and some $\gamma \geq 0$; and $\left(x_L,x_H\right)$ in the red region are optimally implemented by $\gamma = 0$ and some $\beta \geq 0$.}
\label{fig2}
\end{figure}

When the cost function is the entropy cost, it is straightforward to characterize the four regions of $\left\{x_L, x_H\right\}$ pairs. They are depicted in \autoref{fig2}. Here is an \href{https://www.desmos.com/calculator/q3d4h4abxm}{Interactive Link}, where one can adjust the sliders for \(m \equiv \mu\) and \(u \equiv \frac{v_0}{\kappa}\), to see how the optimal contract changes.

\subsubsection{Two States and No Interim IR}

When there is no interim IR constraint, we need only impose $f\left(\mu\right) \geq v_0$ to guarantee that the agent accepts the contract. Defining \[\zeta\left(x_L,x_H\right) \coloneqq \left(1-\mu\right)\left(x_Hc'\left(x_H\right) - c\left(x_H\right)\right) -\mu\left(\left(1-x_L\right)c'\left(x_L\right) + c\left(x_L\right)\right) \text{ ,}\] 
we have
\begin{proposition}\label{nointerimir}
The principal can implement $\left\{x_L,x_H\right\}$ efficiently if and only if $v_0/\kappa \geq \zeta\left(x_L,x_H\right)$. Otherwise, $\gamma = \beta = 0$.
\end{proposition}
It is obvious that an exact analog of Corollary \ref{coreff} holds when there is no interim IR constraint. \textit{Viz.}, any pair of posteriors can be implemented efficiently if the outside option is sufficiently large and the cost of acquiring information $\kappa$ is sufficiently small. Moreover, the more information an agent is asked to acquire, the more difficult it is to implement the distribution efficiently.

\begin{figure}
    \centering
    \includegraphics[scale=.15]{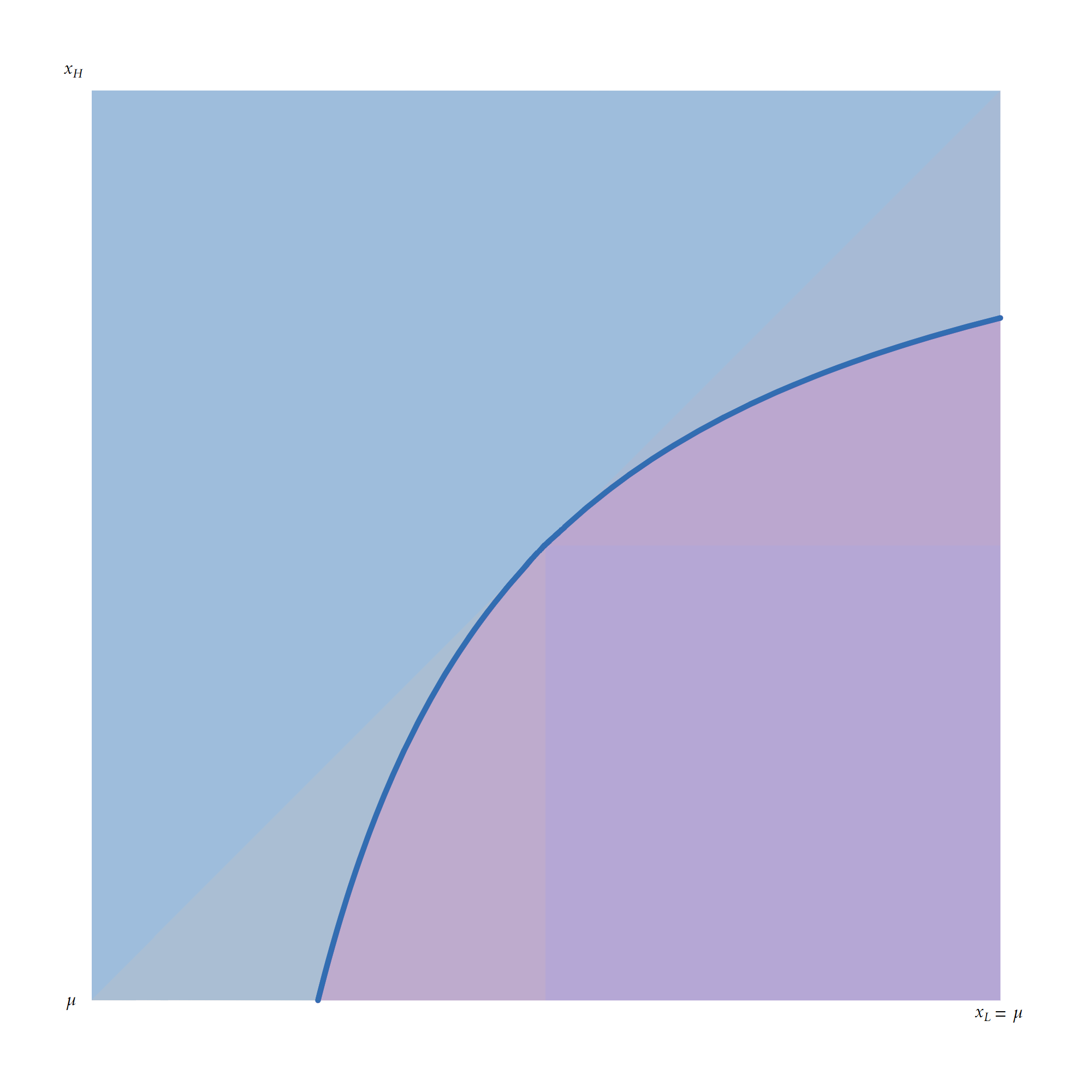}
    \caption{\textbf{Implementation Regions for $\mu = 1/2$ and $v_0/\kappa = \log{\left(1/\left(1-\mu\right)\right)}$ (No Interim IR):} Pairs $\left(x_L,x_H\right)$ in the purple region can be implemented efficiently and $\left(x_L,x_H\right)$ in the blue region are optimally implemented by $\gamma = \beta = 0$.} 
    \label{fig3}
\end{figure}

When the cost function is the entropy cost, it is straightforward to characterize the two regions of $\left\{x_L, x_H\right\}$ pairs. They are depicted in \autoref{fig3}, superimposed over the four regions present when there is an interim participation constraint.

\subsubsection{More Than Two States}
By Lemma \ref{lowest}, for each state $k$ we can find a message $j^*(k)$ that delivers the lowest payment; and by Theorem \ref{dramatic}, to pin down the transfer scheme, it suffices to determine $t^k_{j^*(k)}$ for each state $k$. Thus, there are \(n\) unknowns. When we impose the interim IR constraint, there are \(n\) equations: efficient implementation is equivalent to $f_{\mathcal{H}}(\mu) = v_0$, and the other $n-1$ equations are given by Constraint \ref{icreduced}:
\[t_j^k - t_j^n - \kappa c_k\left(\mathbf{x}_j\right) = -\kappa c_k\left(\mu\right) \text{ ,}\]
where $k = 1, \dots, n-1$ indicates the state, and $j$, which indexes the posterior, is arbitrary. Then the distribution can be efficiently implemented if and only if $t^k_{j^*(k)} \ge 0$ for each $k$; consequently, Proposition \ref{effimp}, Corollary \ref{coreff} \ref{corefi}, Proposition \ref{rncharacterization}, and Proposition \ref{nointerimir} naturally extend to more than two states. 

\section{Discussion}\label{discus}
We conclude by discussing a couple of our assumptions.

\paragraph{Agent has no intrinsic preferences for learning.} We make this assumption to zero in the incentive provision problem when the principal has to delegate information acquisition to an agent who cannot make verifiable reports. To allow the agent to have intrinsic motivation, we assume that the agent's intrinsic value from posterior $\mathbf{x}$ is $\phi(\mathbf{x})$, which is known to both parties. Then it is as if the agent's cost of arriving at posterior $\mathbf{x}$ is $\kappa c(\mathbf{x}) - \phi(\mathbf{x})$, and all of our results survive intact.

\paragraph{Restricted learning.} Our agent is unconstrained in how she learns: she may choose any Bayes-plausible distribution. How might our results change if the agent instead could only choose from some subset thereof? In general, such restrictions must make implementation (of those available distributions) cheaper, as the agent has fewer possible deviations. A corollary of this observation (proved in the supplementary appendix), is that the principal can implement any (feasible) distribution efficiently if the agent is risk neutral and there are no (\emph{ex post}) limited liability constraints.



\bibliography{sample.bib}

\appendix

\section{Omitted Proofs} \label{proofs}

\subsection{Lemma \ref{finducible} Proof}
\begin{proof}
Let $\supp(F) = \{\mathbf{x}_1, \mathbf{x}_2, \dots, \mathbf{x}_{s}\}$, where $s = \left|\supp(F)\right| \le n$. Consider a contract $(M,t)$ where $M = \supp(F)$, and for each $j = 1, \dots, s$,
\[t\left(\mathbf{x}_j, \left.\theta_k \right| \tau\right) = \kappa c\left(\mathbf{x}_{j}\right) - \sum_{i=1}^{n-1} x^{i}_{j} \kappa c_i\left(\mathbf{x}_j\right) + \kappa c_k\left(\mathbf{x}_j\right) + \tau \text{ for all } k = 1, \dots, n-1 \text{ ,}\]
\[t\left(\mathbf{x}_j, \left.\theta_n \right| \tau \right) = \kappa c\left(\mathbf{x}_{j}\right) - \sum_{i=1}^{n-1} x^{i}_{j} \kappa c_i\left(\mathbf{x}_j\right) + \tau \text{ ,}\]
where $x^i_j$ is the $i$-th entry of $\mathbf{x}_j$, $c_i$ is the partial derivative of $c$ with respect to its $i$-th entry, and $\tau$ is a constant that scales the transfers.

For any $m \in M$, we define the agent's \emph{net utility} $N\left(\left.\mathbf{x} \right| m\right)$ as the expected utility of sending message $m$ net of the cost of $\mathbf{x}$: 
\[N\left(\left.\mathbf{x} \right| m \right) = \mathbb{E}_{\mathbf{x}}\left[t\left(m,\theta\right)\right] - \kappa c\left(\mathbf{x}\right) \text{ .}\]
Let $G$ be a distribution over posteriors, and let $\sigma: \Delta (\Theta) \to \Delta(M)$ denote a reporting strategy. Then, the agent's \emph{ex ante} value of choosing $(G, \sigma)$ is given by
\[\Upsilon(G, \sigma) = \sum_{\mathbf{x} \in \supp(G)} \sum_{m \in M} G(\mathbf{x}) \sigma(\left.m\right| \mathbf{x}) N(\left.\mathbf{x} \right| m) \text{ .}\]

We claim that $\left(F, \sigma^*\right)$, where $\sigma^*(\left.\cdot\,\right| \mathbf{x}_j) = \delta_{\mathbf{x}_j}$ is an optimal strategy for the agent.\footnote{\(\delta_{\mathbf{x_j}}\) denotes the degenerate distribution at \(\mathbf{x}_j\); that is, the agent truthfully reports the posterior that he obtained from learning.} By Lemma 1 in \cite{caplin2022rationally}, it suffices to show that, for every $\mathbf{x}_j$, $j = 1, \dots, s$, there exists a $n-1$ dimensional vector $\lambda = (\lambda_1, \dots, \lambda_{s})$ such that
\[N(\left.\mathbf{x} \right| m) - \sum_{i=1}^{n-1} \lambda_i x^i \le N\left(\left.\mathbf{x}_j \right| \mathbf{x}_j\right) - \sum_{i=1}^{n-1} \lambda_i x^i_j\text{ ,}\]
for all $\mathbf{x} \in \Delta (\Theta)$ and $m \in M$. We set $\lambda$ to be the zero vector, so the above inequality reduces to $N(\left.\mathbf{x} \right| m) \le N\left(\left.\mathbf{x}_j \right| \mathbf{x}_j\right)$. We first show that for any fixed $m \in M$, $N(\left.\mathbf{x} \right| m) \le N\left(\left.\mathbf{x}_j \right| m\right)$, and then we show that $N\left(\left.\mathbf{x}_j \right| m\right) \le N\left(\left.\mathbf{x}_j \right| \mathbf{x}_j\right)$. To establish the first inequality, since $c(\mathbf{x})$ is strictly convex, the first-order conditions (FOCs) are sufficient; the FOCs are
\[t\left(m, \left.\theta_i \right| \tau\right) - t\left(m, \left.\theta_n \right| \tau \right) - \kappa c_i\left(\mathbf{x}\right) = \kappa \left(c_i\left(\mathbf{x}_j\right) - c_i\left(\mathbf{x}\right)\right) = 0 \text{ for all } i = 1,\dots,n-1 \text{ ,}\]
clearly setting $\mathbf{x} = \mathbf{x}_j$ makes all of them hold. For the second inequality,
\[N\left(\left.\mathbf{x}_j \right| \mathbf{x}_j\right) - N\left(\left.\mathbf{x}_j \right| m\right) = \kappa \left(c\left(\mathbf{x}_j\right) - c\left(m\right) - \sum_{i=1}^{n-1} \left(x^i_j - m_i\right) c_i(m)\right) \ge 0\text{ ,}\]
where $m_i$ is the $i$-th coordinate of $m$, and the inequality follows from the convexity of $c$. Therefore, $(F,\sigma^*)$ is indeed optimal, and it is direct that the agent's payoff is $\Upsilon(F,\sigma^*) = \tau$. Moreover, there exists $\tau^* < \infty$ large enough, since $c$ is bounded and differentiable on $\inter \Delta (\Theta)$, such that Constraint \ref{icorig} holds. Thus, contract $(M,t)$ implements \(F\). The principal's expected cost is finite since $t\left(\mathbf{x}_j, \left.\theta_k \right| \tau^*\right)$ is finite for all $j, k$.
\end{proof}

\subsection{Proposition \ref{housekeeping} Proof}
\begin{proof}
    Let $\text{ext} \, \mathcal{F}(\mu)$ denote the set of extreme points of $\mathcal{F}(\mu)$. Because $\mathcal{F}(\mu)$ is convex and compact, by Choquet's theorem, for any $G \in \mathcal{F}(\mu)$ there exists a probability measure $\Lambda_G$ that puts probability 1 on $\text{ext} \, \mathcal{F}(\mu)$, and 
    \[\label{representation} \tag{$R$} G = \int_{\text{ext} \, \mathcal{F}(\mu)} H \, \mathrm{d} \Lambda_G(H) \text{ .}\]
    Therefore, any distribution $G$ with support on $\inter \Delta (\Theta)$ can be obtained by randomizing over distributions supported on at most \(n\) affinely-independent points. Then by Lemma \ref{finducible}, $G$ can be implemented at a finite cost by randomizing over contracts we constructed therein. This establishes part \ref{finitecost}. 
    
    For part \ref{wlog}, suppose there exists a contract $(M,t)$ under which the agent chooses $G$, where $|\supp(G)| > n$, and $(G, \hat{\sigma})$ is the induced optimal strategy of the agent. Without loss of generality, $M = \supp(G)$ and $\hat{\sigma}\left(\left.\cdot \right| \mathbf{x}\right) = \delta_{\mathbf{x}}$ for all $\mathbf{x} \in \supp(G)$. Then for every posterior $\mathbf{x} \in \supp(G)$ and every $m \in M$ with $\hat{\sigma}(\left.m\right| \mathbf{x}) > 0$, 
    \[\label{hyperplane} \tag{$H$} N\left(\left.\mathbf{x} \right| m\right) + \sum_{i=1}^{n-1}\left(t\left(m, \theta_i\right) - t\left(m, \theta_n\right) - \kappa c_i\left(\mathbf{x}\right)\right)\left(\tilde{x}_i - x_{i}\right) \ge N\left(\left.\tilde{\mathbf{x}} \right| m'\right)\]
    for all $\tilde{\mathbf{x}} \in \Delta(\Theta)$ and $m' \in M$. By Equation \ref{representation}, for every $F \in \supp(\Lambda_G)$, and every posterior $\mathbf{x}$, $\mathbf{x} \in \supp(G)$. Hence, the strategy $\left(F, \hat{\sigma}\big|_{\supp(F)}\right)$ is also optimal for the agent since Inequality \ref{hyperplane} holds for every $\mathbf{x} \in \supp(F)$ and every $m \in M$ with $\hat{\sigma}\big|_{\supp(F)}(\left.m\right| \mathbf{x}) > 0$. Now it is direct that each $F \in \supp(\Lambda_G)$ can be implemented by the contract $\left(M_F,t_F\right)$ where $M_F = \supp(F)$, and $t_F$ is the restriction of $t$ to $M_F$; thus, $G$ can be implemented at the same cost by randomizing over $\supp(\Lambda_G)$. 
    
    Note that; however, for all $F \in \supp(\Lambda_G)$, $\left(M_F,t_F\right)$ need not be the least costly contract under which the agent chooses \(F\): randomizing over $\supp(\Lambda_G)$ and finding the least costly contract for each \(F\) is at least cheaper than $(M,t)$. Therefore, without loss of generality, the principal only implements distributions with support on at most \(n\) affinely-independent points. This concludes the proof of part \ref{wlog}.
\end{proof}

\subsection{Theorem \ref{dramatic} Proof}

\begin{proof}
The principal wants to implement a distribution \(F\) using some contract $(M,t)$. By part \ref{fci1} of Lemma \ref{fcontrimp}, a necessary condition for implementation is that $\supp(F) = P_{(M,t)}$; this condition holds if and only if the contract is such that the following $s$ expressions
\[\begin{split}&\sum_{k = 1}^{n-1}x_{1}^{k}t_{1}^{k} + \left(1-\sum_{k = 1}^{n-1}x_{1}^{k}\right) t_1^n - \kappa c\left(\mathbf{x}_1\right) + \sum_{k=1}^{n-1}\left(t_1^k - t_1^n - \kappa c_k\left(\mathbf{x}_1\right)\right)\left(x_k - x_1^k\right)\\
&\sum_{k = 1}^{n-1}x_{2}^{k} t_{2}^{k} + \left(1-\sum_{k = 1}^{n-1}x_{2}^{k}\right) t_2^n - \kappa c\left(\mathbf{x}_2\right) + \sum_{k=1}^{n-1}\left(t_2^k - t_2^n - \kappa c_k\left(\mathbf{x}_2\right)\right)\left(x_k - x_2^k\right)\\
&\vdots\\
&\sum_{k = 1}^{n-1}x_{s}^{k}t_{s}^{k} + \left(1-\sum_{k = 1}^{n-1}x_{s}^{k}\right) t_{s}^n - \kappa c\left(\mathbf{x}_{s}\right) + \sum_{k=1}^{n-1}\left(t_{s}^k - t_{s}^n - \kappa c_k\left(\mathbf{x}_{s}\right)\right)\left(x_k - x_{s}^k\right)
\end{split} \text{ ,}\]
define the same hyperplane, where $t^k_j \coloneqq t\left(\mathbf{x}_j,\theta_k\right)$, $x^i_j$ is the $i$-th entry of $\mathbf{x}_j$, and $c_i$ is the partial derivative of $c$ with respect to its $i$-th entry. Accordingly, for all $k = 1, \dots, n-1$ and $i, j = 1, \dots s$
\[t_i^k - t_i^n - \kappa c_k\left(\mathbf{x}_i\right) = t_j^k - t_j^n - \kappa c_k\left(\mathbf{x}_j\right) \quad \text{and} \quad t_i^n = t_j^n + \Xi_{i j} \text{ ,}\]
where $\Xi_{i j}$ is some function of the primitives (but not directly of the $t$s). Combining these two equations, we obtain
\[\Omega^k\left(i,j\right) = \kappa c_k\left(\mathbf{x}_i\right) - \kappa c_k\left(\mathbf{x}_j\right) + \Xi_{i j}, \ \text{for} \ k = 1,\dots,n-1 \ \text{and} \ \Omega^n(i,j) = \Xi_{ij} \text{ .}\]

Accordingly, for each state $k = 1, \dots, n$, once the principal chooses the transfer for one of the messages in state $k$, the transfers for all other messages are automatically pinned down. In other words, the principal has one degree of freedom for each of the states. In every state $k = 1, \dots, n$, and for every $i, j = 1, \dots, s$, we can write $t^k_i = t^k_j + \Omega^k\left(i,j\right)$. \end{proof}

\subsection{Proposition \ref{effinoll} Proof}
\begin{proof}
Let \(F\) be such that $\supp(F) = \{\mathbf{x}_1, \dots, \mathbf{x}_{s}\} \subseteq \inter \Delta (\Theta)$, where $s \le n$. As noted in the main text, there are $n-1$ equations given by Constraint \ref{icreduced}: $t_j^k - t_j^n - \kappa c_k\left(\mathbf{x}_j\right) = -\kappa c_k\left(\mu\right)$ for all $k=1, \dots, n-1$, and efficient implementation requires $f_{\mathcal{H}}(\mu) = v_0$, which can be written as
\[\sum_{k=1}^{n-1} \left(t_j^k - t_j^n - \kappa c_k\left(\mathbf{x}_j\right)\right) \mu_k + t_j^n = Q \text{ ,}\]
where $\mu_k$ is the $k$-th entry of $\mu$, and $Q$ does not depend on $t$'s. To show that \(F\) can be efficiently implemented, it suffices to find a solution of this system of \(n\) equations. Using \ref{icreduced}, the equality above can be reduced to $t^n_j = Q + \sum_{k=1}^{n-1} \kappa \mu_k c_k\left(\mu\right)$; plugging this into the other $n-1$ equations, we get $t^k_j = Q + \sum_{i=1}^{n-1} \kappa \mu_i c_i\left(\mu\right) + \kappa \left(c_k\left(\mathbf{x}_j\right) - c_k\left(\mu\right)\right)$ for each $k=1, \dots, n$. We have thus found a solution. Because \(F\) is an arbitrary distribution over posteriors supported on at most \(n\) points, the principal can implement any distribution $G$ with $\supp(G) \subseteq \inter \Delta (\Theta)$ efficiently by randomizing \textit{ex ante}. 
\end{proof}

\subsection{Proposition \ref{riskaverse} Proof}
\begin{proof}
    Suppose first that the agent can exit \emph{ex interim}. Because $c$ is strictly convex, $v_0 - c(\mathbf{x})$ is strictly concave. By Observation \ref{ratangency}, $f_\mathcal{H}\left(\mathbf{x}\right)$ is tangent to $v_0 - c(\mathbf{x})$ at $\mathbf{x}^*$. Thus, $\mathbf{x}^* \ne \mu$ implies that $f_{\mathcal{H}}(\mu) > v_0$, and hence the agent gets strictly positive rents. When the agent cannot exit \emph{ex interim}, the fact that he gets zero rents is almost immediate: if $f_{\mathcal{H}}(\mu) > v_0$, because there is no limited liability, the transfer can be lowered by some small $\varepsilon > 0$. 
    
    Let \(F\) be a nondegenerate distribution with $\supp{(F)} = \{\mathbf{x}_1, \dots, \mathbf{x}_s\}$, where $\mathbf{x}_i \ne \mathbf{x}_j$ for all $i,j = 1, \dots, s$ with $i \ne j$. Suppose to the contrary that \(F\) can be efficiently implemented. To simplify notation, let $x_j^n := 1 - \sum_{k=1}^{n-1} x_j^k$ for each $j=1, \dots, s$. The principal wishes to minimize $\sum_{j=1}^{s} \sum_{k=1}^{n} p_j x_j^k v^{-1}\left(t^k_j\right)$. Because \(F\) can be efficiently implemented, 
    \[\sum_{j=1}^{s} \sum_{k=1}^{n} p_j x_j^k v^{-1}\left(t^k_j\right) = v^{-1}\left(C(F) + v_0\right) = v^{-1}\left(\sum_{j=1}^{s} \sum_{k=1}^{n} p_j x_j^k t^k_j\right) \text{ .}\]
    Because $v$ is strictly concave, $v^{-1}$ is strictly convex; Jensen's inequality then implies $t^k_j = \tilde{t}$ for all $k=1,\dots,n$ and $j = 1, \dots, s$. But then since $c$ is strictly convex, learning according to the degenerate distribution is uniquely optimal to the agent, and hence the contract with constant transfer cannot implement \(F\). A contradiction. 
\end{proof}

\subsection{Lemma \ref{lowest} Proof}
\begin{proof}
Fix any state $k$ and an arbitrary message, say $s$. Define $N(k) = \left\{i : \Omega^{k}\left(i,s\right) < 0 \right\}$. If $N(k) = \varnothing$, let $j^{*}(k) = s$; then since $t^k_i = t^k_{j^*(k)} + \Omega^k\left(i,s\right)$, we have $t^k_{j^*(k)} \le t^k_i$ for all $i=1,\dots,s$. Otherwise, let $j^{*}(k)$ be an arbitrary selection of $\argmin_{j \in N(k)} \Omega^{k}(j,s)$. Optimal learning requires, for any $i$, $t^k_i = t^k_{s} + \Omega^k\left(i,s\right)$ and $t^k_{j^*(k)} = t^k_{s} + \Omega^k\left(j^*(k),s\right)$, which implies $t^k_{j^*(k)} - t^k_i = \Omega^k\left(j^*(k),s\right) - \Omega^k\left(i,s\right) \le 0$.
Again, $t^k_{j^*(k)} \le t^k_i$ for all $i=1,\dots,s$. 
\end{proof}

\subsection{Proposition \ref{llbinds} Proof}
\begin{proof}
By Lemma \ref{lowest}, for every state $k = 1, \dots, n$, there exists $j^*(k)$ such that $t_{j^{*}(k)}^{k} \le t_i^k$ for all $i=1,\ldots,s$. Then by setting $t_{j^{*}(k)}^{k} = 0$, the agent's honesty is not affected, and the limited liability constraints are satisfied. For every $i \ne j^{*}(k)$, we have
\[t^{k}_{i} = \Omega^{k}(i,j^{*}(k)) = \kappa c_k\left(\mathbf{x}_i\right) - \kappa  c_k\left(\mathbf{x}_{j^{*}(k)}\right) + \Xi_{i j^{*}(k)}\]
for each $k = 1, \dots, n-1$; and $t^{n}_{i} = \Xi_{i j^{*}(n)}$ for $i \ne j^{*}(n)$.
\end{proof}

\subsection{Proposition \ref{effimp} Proof}
\begin{proof}
Without loss of generality $\alpha \coloneqq t_{1}^{1} \geq t_{2}^{1} \eqqcolon \gamma$; and $\delta \coloneqq t_2^2 \geq t_1^2 \eqqcolon \beta$. In this case, it is convenient to write down the agent's value function:
\[W\left(x\right) = \begin{cases}
\alpha \left(1-x\right) + \beta x - \kappa c\left(x\right), \quad &\text{if} \quad 0 \leq x \leq \frac{\alpha - \gamma}{\alpha - \gamma + \delta - \beta}\\
\gamma \left(1-x\right) + \delta x - \kappa c\left(x\right), \quad &\text{if} \quad \frac{\alpha - \gamma}{\alpha - \gamma + \delta - \beta} \leq x \leq 1\\
\end{cases} \text{ .}\]
Consequently, the equations that pin down the agent's optimal learning simplify to
\[\kappa \left(c'\left(x_H\right) - c'\left(x_L\right)\right) = A + B \quad \text{and} \quad A + \kappa \left(c\left(x_H\right) - c\left(x_L\right)\right) = \kappa \left(c'\left(x_H\right) x_H - c'\left(x_L\right) x_L\right) \text{ ,}\]
where $A \coloneqq \alpha - \gamma \geq 0$, $B \coloneqq \delta - \beta \geq 0$. Because $c$ is strictly convex, both $A$ and $B$ are strictly positive if $x_L < \mu < x_H$, and zero if $x_L = x_H = \mu$. Furthermore, the concavifying line is
\[\label{cline} \tag{$\star$} f\left(x\right) = \left(\beta - \gamma - A - \kappa c'\left(x_L\right)\right) x + \gamma + A - \kappa \left(c\left(x_L\right) - x_L c'\left(x_L\right)\right) \text{ .}\] 
The principal chooses $\gamma$ and $\beta$ in order to maximize
\[- \gamma \left(1-\mu\right) - \beta \mu - p x_H B - \left(1-p\right) \left(1-x_L\right) A \text{ ,}\]
where $p = (\mu - x_L)/(x_H - x_L)$ is the (unconditional) probability that posterior $x_H$ realizes,
subject to limited liability: $\beta, \gamma \ge 0$, and \[\label{icoo}\tag{$IR$-\(v_{0}\)}f\left(x\right) \geq v_0 - \kappa c\left(x\right) \quad \text{for all} \quad x \in \left[0,1\right] \text{ ,}\]
where \(F\) is given in Equation \ref{cline}. By construction, the agent cannot deviate profitably by learning differently \textit{and} reporting to the principal. Constraint \ref{icoo} ensures that the agent cannot deviate profitably by learning differently and taking his outside option.

Using the concavifying line (\ref{cline}), $\left\{x_L,x_H\right\}$ can be implemented efficiently if and only if 
\begin{enumerate}[label={(\roman*)},noitemsep,topsep=0pt]
    \item\label{1i} $\left(\beta - \gamma - A - \kappa c'\left(x_L\right)\right) \mu + \gamma + A - \kappa \left(c\left(x_L\right) - x_L c'\left(x_L\right)\right) = v_0$; and
    \item\label{2i} $\beta - \gamma - A - \kappa c'\left(x_L\right) = - \kappa c'(\mu)$; and
    \item\label{3i} $\beta, \gamma \ge 0$.
\end{enumerate}
From \ref{1i} and \ref{2i},
\[\gamma = v_0+ \kappa c'(\mu) \mu-A-\kappa\left(c'\left(x_{L}\right) x_{L}-c\left(x_{L}\right)\right) = v_0+ \kappa c'(\mu) \mu - \kappa\left(c'\left(x_{H}\right) x_{H}-c\left(x_{H}\right)\right) \text{ ,}\]
and
\[\beta = v_0-\kappa (1-\mu) c'(\mu)+\kappa\left(1-x_{L}\right) c'\left(x_{L}\right)+ \kappa c\left(x_{L}\right) \text{ .}\]
\ref{3i} requires $v_0/\kappa \ge \eta\left(x_L,x_H\right)$, as stated in the result.
\end{proof}

\subsection{Proposition \ref{rncharacterization} Proof}
\begin{proof}
\ref{rnci} is a consequence of Proposition \ref{effimp}. Suppose that $v_0/\kappa < \eta\left(x_L,x_H\right)$ so that efficient implementation is infeasible. Recall that $P$ wants to maximize $-\gamma \left(1-\mu\right) - \beta \mu$. Thus, if $\gamma = \beta = 0$ is implementable, they are obviously optimal. Substituting them into the concavifying line (\ref{cline}) we get 
\[h\left(x\right) = -\left(A + \kappa c'\left(x_L\right)\right) x + A - \kappa \left(c\left(x_L\right) - x_L c'\left(x_L\right)\right) \text{ .}\] 
We need to check for which values of $x_L$ and $x_H$ $h$ lies above $v_0 - \kappa c\left(x\right)$. To that end, we define function $g\left(x\right) \coloneqq h\left(x\right) - v_0 + \kappa c\left(x\right)$. Then,
\[g'\left(x\right) = -\left(A + \kappa c'\left(x_L\right)\right) + \kappa c'\left(x\right) \text{ ,}\]
and observe that $g$ is strictly convex in $x$. Evidently, $g'\left(0\right) < 0$, so \(F\) is either minimized at $x^{\circ} = x^{\circ}\left(x_L,x_H\right)$, implicitly defined as $g'\left(x^\circ\right) = 0$ (if such an $x \leq 1$ exists) or $x = 1$. Define $x^{\dagger} \coloneqq \min\left\{x^{\circ}, 1\right\}$. Thus, $\gamma = \beta = 0$ is optimal if and only if $g\left(x^{\dagger}\right) \geq 0$. Note that there is a knife-edge case where $v_0/\kappa = \eta \left(x_L,x_H\right)$, $x^{\dagger} = \mu$, and $\beta = \gamma = 0$ (and the first-best is attained). This is the only way for all three constraints to bind.

Can we have one of the non-negativity constraints bind, $\gamma = 0$, say; and the other constraints all be slack, i.e., $\beta > 0$ and $f\left(x\right) > v_0 - \kappa c\left(x\right)$ for all $x \in \left[0,1\right]$? No: otherwise the principal could decrease $\beta$ by a sufficiently small $\varepsilon > 0$, strictly increasing her payoff and still leaving Constraint \ref{icoo} satisfied. This yields \ref{rncii} and \ref{rnciii} of the result.
\end{proof}

\subsection{Proposition \ref{nointerimir} Proof}
\begin{proof}
Simply rearrange the inequality $f\left(\mu\right) \geq v_0$ to get 
\[\gamma \geq \frac{v_0 - \kappa \left(\left(1-\mu\right)\left(x_Hc'\left(x_H\right) - c\left(x_H\right)\right) -\mu\left(\left(1-x_L\right)c'\left(x_H\right) + c\left(x_L\right)\right)\right)}{1-\mu} - \frac{\mu}{1-\mu}\beta \text{ ,}\]
then set $\beta = 0$ and solve for when the right-hand side of this inequality is positive.\end{proof}

\newpage

\section{Supplemental Appendix}

\subsection{When the Agent ``Skips Town'' (An Example)}\label{skiptown}

The purpose of this subsection is to illustrate how the principal may wish to have the agent take his outside option with positive probability if such an exit does not hurt the principal (too much). First, a minor remark that justifies one of our assumptions in the text:

\begin{remark}\label{noexit}
If the principal's payoff from the agent taking his outside option is $< -v_0$, there is no optimal contract in which the agent takes his outside option with positive probability. 
\end{remark}
\begin{proof}
The result is nearly immediate: suppose for the sake of contradiction there is an optimal contract in which the agent takes his outside option with strictly positive probability. Replace that contract with an identical one with one exception: now the principal offers a constant payout of $v_0$ to the agent for reporting the previous ``null message belief.''  The agent's incentives are unaffected and the principal's payoff is strictly higher, contradicting the original contract's purported optimality. \end{proof}

Next we will illustrate that if the disutility incurred by the principal when the agent takes his outside option is not too low, the principal may prefer that the agent not have an interim participation constraint,\footnote{A moment's reflection reveals that this statement is either incorrect or misleading: introducing an additional constraint in an optimization problem can never strictly increase the optimal payoff. What engenders the (potential) improvement here is that the addition of the interim IR constraint is also the addition of new contractual possibilities (contracts in which the agent exits the relationship after learning).} which allows the principal to write contracts in which the agent exits the relationship with positive probability.

\begin{figure}
\centering
\begin{subfigure}{.85\textwidth}
  \centering
  \includegraphics[scale=.2]{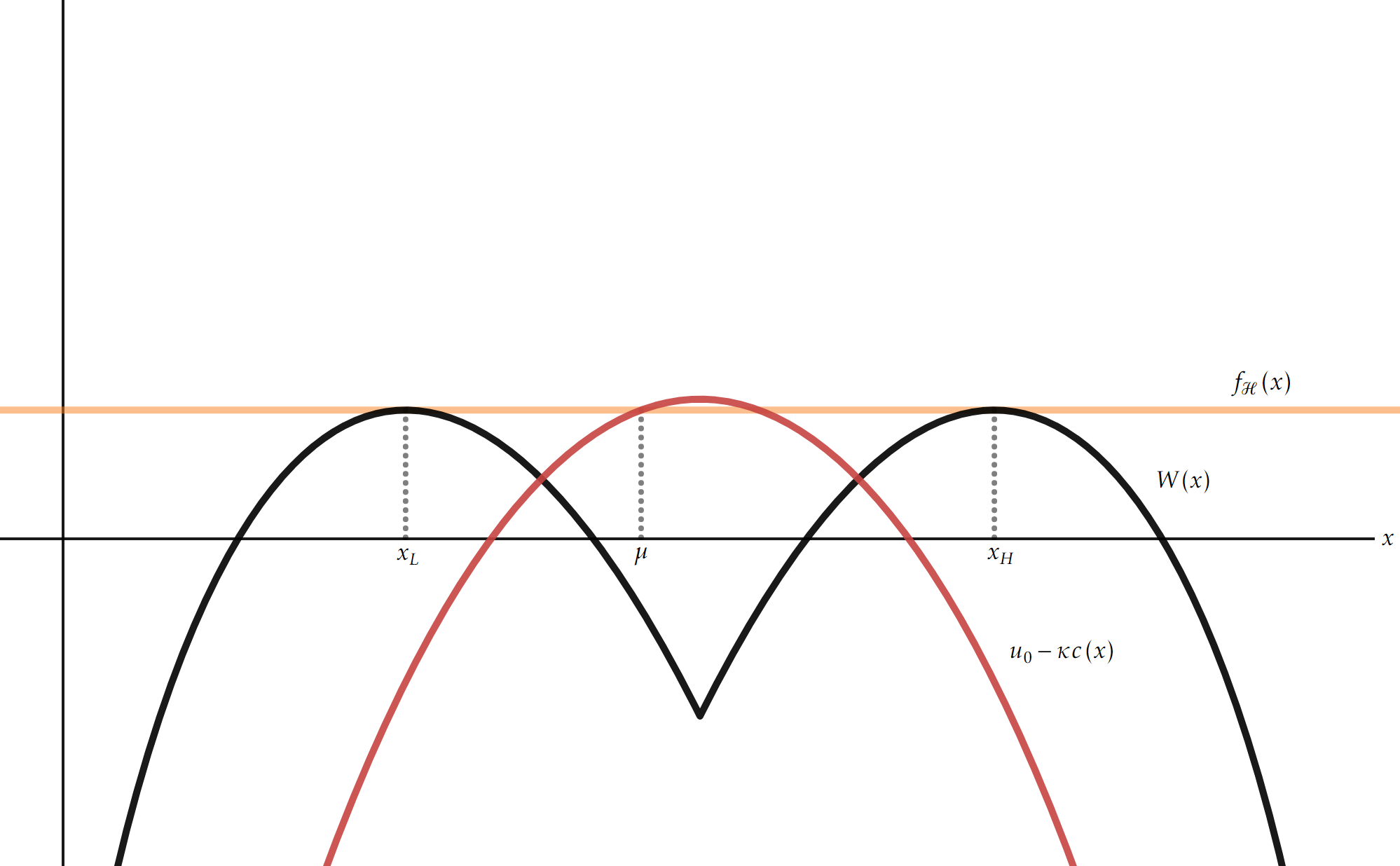}
  \caption{An optimal contract (the STP contract) when the agent may not exit \textit{ex interim}}
  \label{figsuba1}
\end{subfigure}%
\par
\bigskip
\par
\bigskip
\par
\begin{subfigure}{.85\textwidth}
  \centering
  \includegraphics[scale=.2]{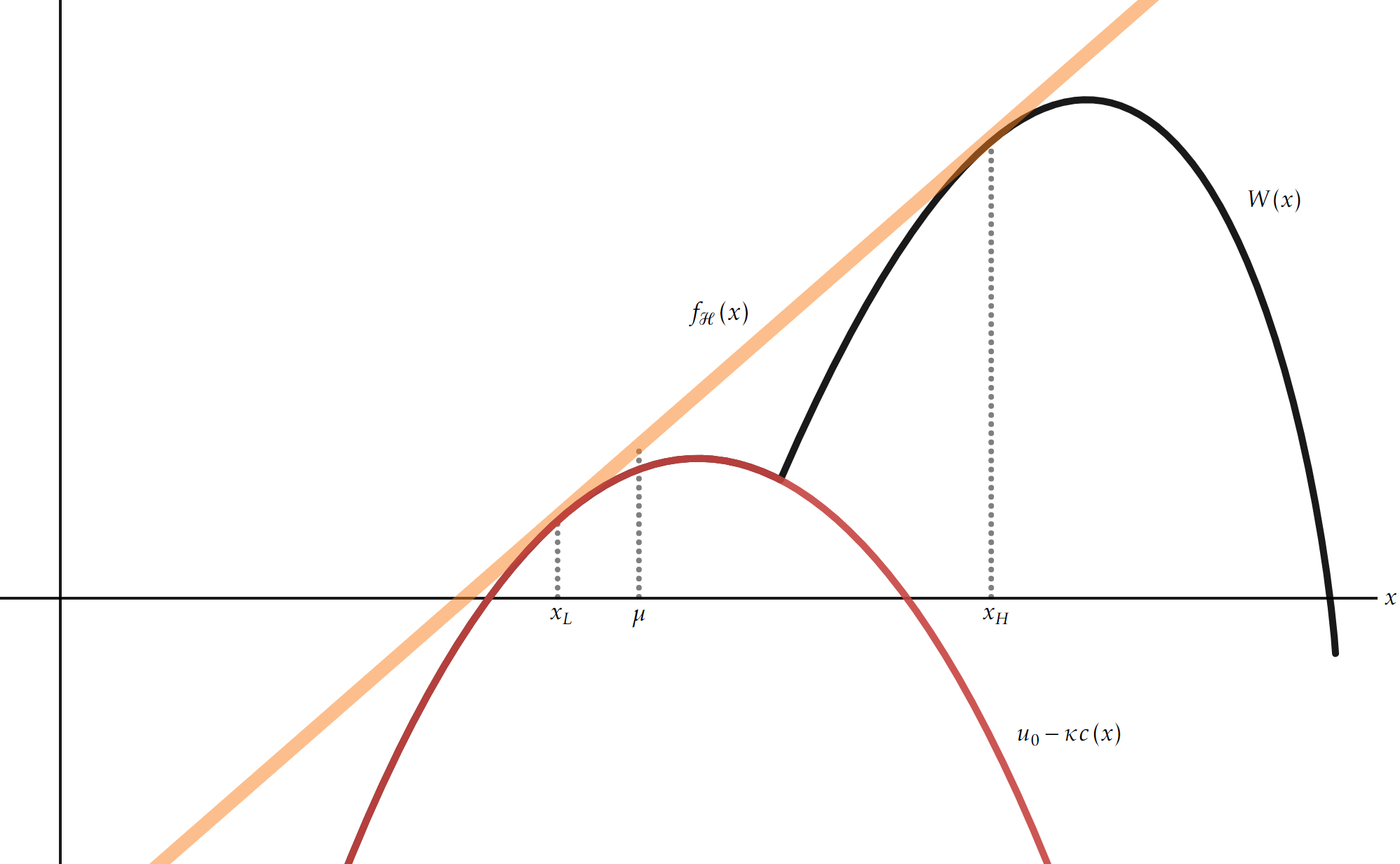}
  \caption{A contract yielding a strict Pareto improvement to any ``no exit'' contract}
  \label{figsuba2}
\end{subfigure}
\caption{In both \ref{figsuba1} and \ref{figsuba2}, $\mu \approx .45$. In the former, $x_L \approx .27$, $x_H \approx .73$, and the concavifying line $f_\mathcal{H}(x) = v_0 = .05$. In the latter, $x_L \approx .39$, $x_H \approx .73$ and $f_\mathcal{H}(x)$ is approximately $.44x-.14$.}
\label{figa1}
\end{figure}

Consider the following example, which is illustrated in Figure \ref{figa1}. The state is binary, $\Theta = \left\{0,1\right\}$, and $\mathbb{P}\left(1\right) = \left(3+2e\right)/\left(5+5e\right) \approx .45$. The principal's decision problem is a simple ``match the state'' task: she has two actions $a \in \left\{0,1\right\}$ and obtains a payoff of $1$ if $a = \theta$ and $0$ otherwise. The agent's cost of acquiring information is the entropy cost, the cost parameter is $\kappa = 1$ for simplicity, and she has an outside option of $.05$. The principal suffers no disutility if the agent takes her outside option.

If the principal controlled information herself, her distribution over posteriors would have support $\left\{1/\left(1+e\right), e/\left(1+e\right)\right\} \approx \left\{.27 , .73\right\} $. As we note in Proposition 5.1, the principal can still implement this distribution efficiently even with the interim IR constraint. This distribution is optimal, therefore, when the agent may not exit the relationship after learning, which yields the principal an approximate payoff of $.62-.05 = .57$ (the payoff in the decision problem net of the information cost minus the agent's outside option). The value function for the agent induced by the STP contract, which is optimal when there is no interim participation constraint, and the agent's resulting optimal learning are depicted in Figure \ref{figsuba1}.

Now let us allow the agent to exit the relationship after learning and construct a contract that strictly improves the principal's payoff. The contract consists of a single message, $.73$, and a transfer from sending this message of $2/3$ in state $1$ and $-3/4$ in state $0$. The agent's optimal learning now has (approximate) support $\left\{.39,.73\right\}$, after which she exits the relationship, taking her outside option, or sends the single offered message, respectively. This yields the principal a payoff of approximately $.58$, a strict improvement. Moreover, the agent obtains strictly positive surplus as well, approximately $.06$. The value function for the agent induced by this contract as well as the agent's optimal learning are depicted in Figure \ref{figsuba2}.

This contract is not optimal for the principal, but it serves its purpose: the optimal contract in this example must be one in which the agent exits the relationship with positive probability. Nor does the agent exit the relationship with probability one (for this would correspond to him acquiring no information), and so in the optimal contract the agent acquires strictly positive surplus. Thus, allowing an interim exit may engender a strict Pareto improvement.

That enabling an agent to exit the relationship may strictly improve welfare is a consequence of the special kind of output the agent is asked to produce in our setup. Namely, he is asked to provide the principal with information: the informational content of an agent's action (on path) is the same regardless of whether it is conveyed via a message or the agent's resignation. One natural interpretation for the cost an agent's exit imposes on the principal is that it is the cost of finding a new advisor. Consequently, if it is relatively cheap to do so, the principal prefers this to compensating the advisor.\footnote{One interpretation of this observation is as rationalizing the phenomenon of ``shooting the messenger'' or blaming the bearer of bad news--``...Yet the first bringer of unwelcome news, Hath but a losing office...'' (Henry IV, pt. II) Rather than inform the principal himself (and suffer her wrath), the agent prefers to skip town, which nevertheless informs the principal about the situation.}

\subsection{Salvage Value}

Our framework can also easily be adapted to allow the agent to have a ``salvage'' value for a posterior (upper semicontinuous function) $p\left(\mathbf{x}\right)$, from exiting the relationship after obtaining posterior $\mathbf{x}$. This corresponds, for instance, to a scenario in which the agent is able to (credibly) sell information to a third party. In this case, incentive compatibility is left unchanged, but the participation constraint becomes
\[\label{icnew}\tag{$IR-p$} f_{\mathcal{H}}(\mathbf{x}) \ge p\left(\mathbf{x}\right) - \kappa c(\mathbf{x}) \quad \text{for all} \quad \mathbf{x} \in \Delta(\Theta) \text{ .}\]
An analog of Lemma 3.1 is immediate:
\begin{lemma} \label{fcontrimpnew}
    A contract $(M,t)$ implements distribution $F$ if and only if
    \begin{enumerate}[label={(\roman*)},noitemsep,topsep=0pt]
        \item $\supp(F) = P_{(M,t)}$; and 
        \item Constraint \ref{icnew} holds; and
        \item if there is limited liability, $t(m, \theta) \ge 0$ for all $\theta \in \Theta$ and $m \in M$.
    \end{enumerate}
\end{lemma}
Given this, it is easy to see that the other results of the paper go through with the outside option function $v_0 - \kappa c$ replaced with the concavification of $p - \kappa c$.

\subsection{Efficient Implementation With a Restricted Set of Distributions}

In principle, the agent could be further constrained in how she learns: perhaps $F$ must be chosen not from the set of Bayes-plausible distributions $\mathcal{F}_{\mu}$ but from some (weak-$*$) compact subset $\mathcal{P}$ of the Bayes' plausible distributions with finite support.
One example of this is if there are just two states, $\Theta = \left\{0,1\right\}$, the prior is $\frac{1}{2}$, and the agent has access to collection of binary signals $\pi_{\alpha}\left(\left.1\right| 1\right) = \pi_{\alpha}\left(\left.0\right| 0\right) = \alpha \in \left[\frac{1}{2}, 1\right]$. In this case $\mathcal{P}$ is the collection of binary distributions with support $\left\{1-\alpha, \alpha\right\}_{\alpha \in \left[\frac{1}{2},1\right]}$ and mean $\frac{1}{2}$.

We further assume that the cost functional restricted to subset $\mathcal{P}$ is posterior separable. In the parametrized binary experiment example, any strictly convex, twice continuously differentiable function $ c\colon \left[\frac{1}{2},1\right] \to \mathbb{R}_{+}$ with $c\left(\frac{1}{2}\right) = 0$ would do; e.g., $c\left(\alpha\right) = \kappa \left(\frac{\alpha^2}{2} - \frac{1}{8}\right)$ (with $\kappa > 0$).

Even though the set of distributions available to the agent is limited, the principal is nevertheless unconstrained by the agency problem. Namely, an analog of Proposition 5.1 holds:
\begin{proposition} \label{effinolllimit}
    If the agent is risk neutral and not protected by limited liability, every (feasible) distribution $F \in \mathcal{P}$ with $\supp(F) \subseteq \inter \Delta (\Theta)$ can be implemented efficiently.
\end{proposition}
\begin{proof}
Fix any $F \in \mathcal{P}$ with $\supp(F) \subseteq \inter \Delta (\Theta)$. As we note in Corollary 4.2 (and which is a consequence of Choquet's theorem), this $F$ can by obtained by randomizing over Bayes-plausible distributions each with support on at most $n$ points. This collection of distributions is $\left(F_i\right)_{i \in I}$. By Proposition 5.1, each $F_i$ can be implemented efficiently. Moreover, efficiency is synonymous with the tangency of the corresponding hyperplanes $f_{\mathcal{H},i}$ with the outside option curve $v_0 - \kappa c$ at the prior. However, this hyperplane is unique so each $f_{\mathcal{H}, i}$ must equal some common $f_{\mathcal{H}}$. Furthermore, for each contract $\left(M_i, t_i\right)_{i \in I}$, the other part of Lemma 3.1 also must hold: $\supp F_i = P\left(M_i, t_i\right)$ for all $i$. Accordingly, the contract $\left(M, t\right)$--defined as $M = \cup_{i \in I}M_i$ and $t\left(m, \theta\right) = t_i\left(m, \theta\right)$ for all $m \in M_i$ for all $i \in I$--implements $F \in \mathcal{P}$ efficiently.
\end{proof}


\subsection{An Analog of Proposition 5.1 for a Prior with a Density}

Now let us derive an analog of Proposition 5.1 when there are uncountably many states and the prior admits a density.\footnote{The idea for this sort of moment-based problem originates from a joint project of the first author with Andreas Kleiner, Benny Moldovanu, and Philipp Strack (\cite{kleiner2023extreme}), whom we thank.} Suppose the state $\theta$ is distributed on the unit interval according to absolutely continuous cdf $F$. Given any distribution, $G$, we stipulate that the principal's utility function is a convex function of the first $z$ (non-centered) moments \[m_1 = \int_{0}^{1}xdG\left(x\right), \ m_2 = \int_{0}^{1}x^2dG\left(x\right),\dots,m_z = \int_{0}^{1}x^zdG\left(x\right) \text{ .}\]
We write $V\left(m_1, m_2, \dots, m_z\right)$. In addition, the agent's cost of acquiring information is also a convex function of the first $z$ non-centered moments $\kappa c\left(m_1, m_2, \dots, m_z\right)$ (where $\kappa > 0$ is a scaling parameter). 

Associated with the prior $F$ is a joint distribution over $\left(x, x^2, \dots, x^z\right)$, $\hat{F}$, and so the principal's problem--should she control information acquisition herself--is
\[\max_{H \in \mathcal{F}\left(\hat{F}\right)}\int \left(V-\kappa c\right)dH \text{ ,}\]
where $\mathcal{F}\left(\hat{F}\right)$ denotes the set of fusions of $\hat{F}$.\footnote{\cite{kleiner2023extreme} explore this problem in detail.} For simplicity we also assume that the principal also has just $t < \infty$ actions, which ensures that the optimal fusion has support on at most $t$ points. Then,
\begin{proposition}\label{dilation}
If the agent is risk neutral and not protected by limited liability, every distribution $H \in \mathcal{F}\left(\hat{F}\right)$ with support on $t$ or fewer points can be implemented efficiently.
\end{proposition}
\begin{proof}
By definition, distribution $H \in \mathcal{F}\left(\hat{F}\right)$ and has support on $t$ or fewer points in the $z$-dimensional unit hypercube. Denote by $\mu \coloneqq \left(\mu_1, \dots, \mu_z\right)$ the barycenter of measure $\hat{F}$, and let $f_{\mu}$ denote the hyperplane tangent to $u_0 - \kappa c\left(m\right)$ at $\mu$. The principal contracts the contract as follows: for each posterior $m'$ in support of $H$, the contract is such that the agent's payoff gross of the information acquisition cost is of the form
\[\tau_{m'}\left(m\right) = \alpha_{m', 1} m_1 + \alpha_{m', 2} m_2 + \dots + \alpha_{m', z} m_z + \beta_{m'} \text{ ,}\]
where each $\alpha_{m', i}$ and $\beta_{m'}$ are scalars;\footnote{To elaborate, for each $m' \in \supp(H)$, the transfer is given by
\[t\left(m', \theta\right) = \alpha_{m', 1} \theta + \alpha_{m', 2} \theta^2 + \dots + \alpha_{m', z} \theta^z + \beta_{m'} \text{ ,}\]
and hence if the agent receives some signal $\psi$ and reports $m'$, his expected payoff gross of the information acquisition cost is
\[\int_{0}^{1} t(\theta) \, dG_{\psi}(\theta) = \alpha_{m', 1} \int_{0}^{1} \theta \, dG_{\psi}(\theta) + \alpha_{m', 2} \int_{0}^{1} \theta^2 \, dG_{\psi}(\theta) + \dots + \alpha_{m', z} \int_{0}^{1} \theta^z \, dG_{\psi}(\theta) + \beta_{m'} \text{ ,}\]
where $G_{\psi}$ is the distribution over states conditional on the agent's signal. The right-hand side of the equation above is just $\tau_{m'}(m)$.} and such that $\tau_{m'}\left(m\right) - \kappa c\left(m\right)$ is tangent to $f_{\mu}$ at $m'$.

Evidently, given this contract it is optimal for the agent to acquire distribution $H$ in the relaxed problem in which she simply chooses a distribution that is a dilation of the prior $\mu$, and therefore it must be optimal for the agent to acquire distribution $H$ in her fusion problem (and by construction $H \in \mathcal{F}\left(\hat{F}\right)$). Moreover, the agent obtains zero rents.
\end{proof}
Note that this results holds regardless of whether there is an interim participation constraint--if no such constraint exists, the principal can offer the STP contract, and if not (as above), the STP does not satisfy the interim IR constraint generically.

We can also illustrate this result with the following example. The prior is the uniform distribution on the unit interval. The principal has just two actions, and the payoff in the principal's decision problem can be written as a function of the posterior's first moment. Specifically, $V\left(m\right) = \max{\left\{1-m, m\right\}}$. The agent's cost of acquiring information can also be written as a function of the posterior's first moment: $c\left(m\right) = \left(m-1/2\right)^2$. The agent has an outside option of $0$. The principal can implement any (feasible) pair of posterior first moments $m_L$ and $m_H$ (with $m_L < 1/2 < m_H$) by offering the contract \[\left(m_L ; \left(\kappa\left(2m_L-1\right), \frac{\kappa}{4}\left(1-4m_L^2\right)\right)\right), \ \left(m_H ; \left(\kappa\left(2m_H-1\right), \frac{\kappa}{4}\left(1-4m_H^2\right)\right)\right) \text{ ,}\]
where each element of the contract is of the form $\left(m'; \left(\alpha_{m'}, \beta_{m'}\right)\right)$.

\begin{figure}
    \centering
    \includegraphics[scale=.15]{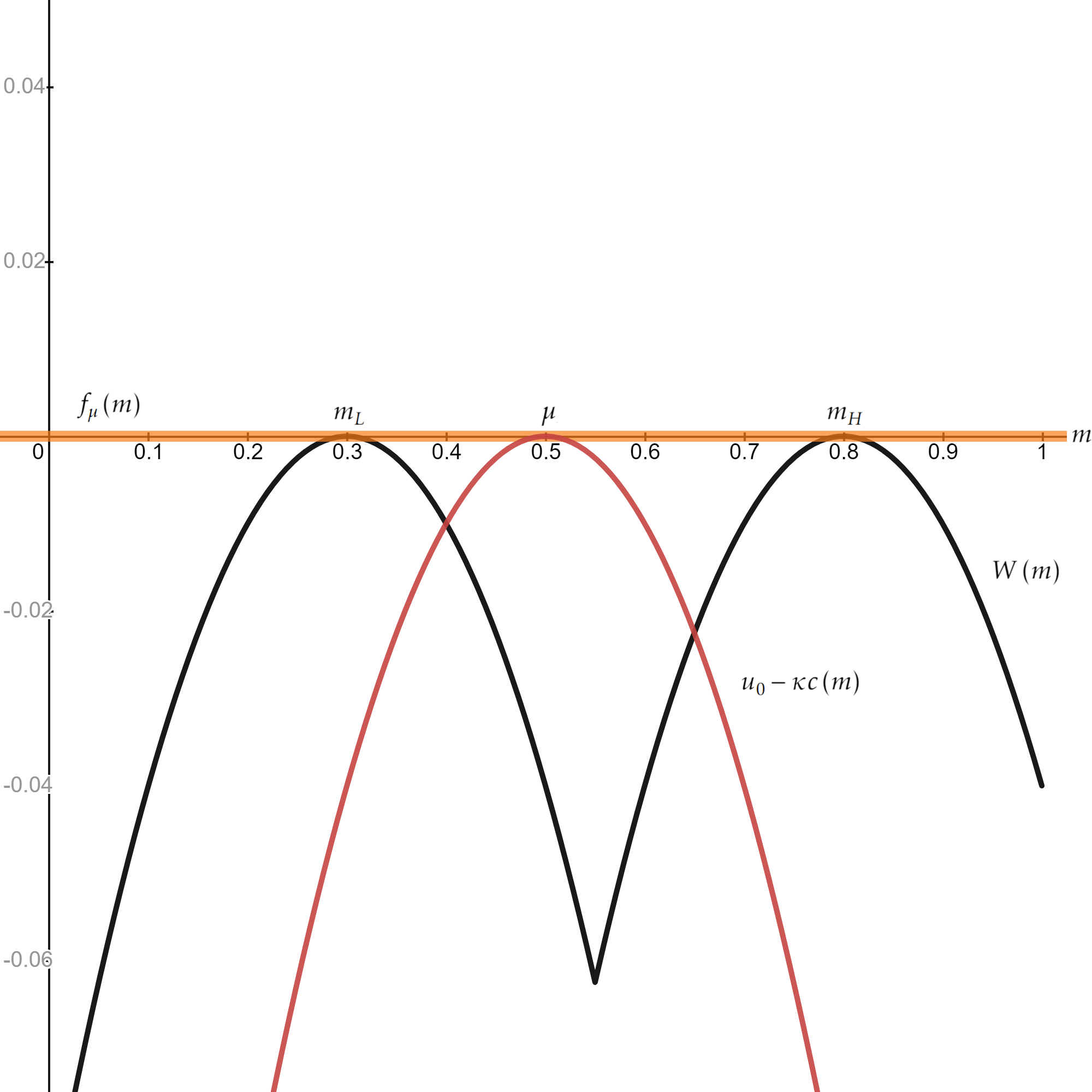}
    \caption{A ``moment acquisition'' example.}
    \label{figapp}
\end{figure}

Figure \ref{figapp} illustrates the induced decision problem for the agent, $W\left(m\right)$ that implements the pair $m_L = .3$ and $m_H = .8$ when $\kappa = 1$.

\subsection{Interim IR in the Canonical Problem (Section 5.1.1.)}\label{interimircan}

Consider the basic moral hazard setting with no limited liability and risk neutral principal and agent. For simplicity the agent chooses effort $a \in \left[0,1\right]$ and her output, $x$, takes values in $\left[0,1\right]$. Now, however, the agent's output is private.  After it realizes she may choose whether to turn in the output and receive the promised remuneration $w\left(x\right)$ or exit the relationship to take her outside option $v_0$. As is standard in the literature, the agent's cost of effort $c(\cdot)$ is strictly increasing and strictly convex, with $c(0) = 0$. Output takes values, $x$, in the unit interval. The family of conditional densities of output realizations $f\left(\left.x \right| a\right)$ has full support for each $a \in [0,1]$ and satisfies the MLRP. Let $\mathcal{X} \subseteq \left[0,1\right]$ denote the subset of agents that the principal wants to report, and $a^*$ is the desired effort level. Both the principal and the agent are risk neutral. 

The result:
\begin{proposition}
Unless the principal is observing the lowest possible effort ($a = 0$) and/or the agent is almost never turning in her output ($\mathcal{X}$ has Lebesgue measure $0$), the agent gets strictly positive rents.
\end{proposition}
\begin{proof}

The principal solves
\[\min_{w\left(\cdot\right)} \int_{\mathcal{X}} w\left(x\right)f\left(\left.x\right|a^{*}\right)dx \text{ ,}\]
subject to
\[\label{canir}\tag{$IR$} \int_{\mathcal{X}} w\left(x\right)f\left(\left.x\right|a^{*}\right)dx + v_0 \int_{\left[0,1\right] \setminus \mathcal{X}} f\left(\left.x\right|a^{*}\right)dx - c\left(a^*\right) \geq v_0 \text{ ,}\]
\[\label{canic}\tag{$IC$} a^{*} \in \argmax_{a}\left\{\int_{\mathcal{X}} w\left(x\right)f\left(\left.x\right|a\right)dx + v_0 \int_{\left[0,1\right] \setminus \mathcal{X}} f\left(\left.x\right|a\right)dx - c\left(a\right)\right\} \text{ ,}\]
and
\[\tag{$IIR$}\label{caniir} w(x) \geq v_0 \ \text{for all} \ x \in \mathcal{X} \text{ .}\]
Observe that unless the principal is implementing $a = 0$ and or $\mathcal{X}$ has (Lebesgue) measure 0, $w\left(x\right) > v_0$ for a positive measure subset of $\mathcal{X}$. Thus, if the principal implements a nontrivial outcome $a > 0$ and $\mathcal{X}$ has strictly positive measure, by \ref{canic} and \ref{caniir},
\[\int_{\mathcal{X}} w\left(x\right)f\left(\left.x\right|a^{*}\right)dx + v_0 \int_{\left[0,1\right] \setminus \mathcal{X}} f\left(\left.x\right|a^{*}\right) dx - c\left(a^*\right) > v_0\int_{\mathcal{X}} f\left(\left.x\right|0\right)dx + v_0 \int_{\left[0,1\right] \setminus \mathcal{X}} f\left(\left.x\right|0\right)dx =v_0 \text{,}\]
and hence \ref{canir} holds strictly. Consequently, the agent gets strictly positive rents. \end{proof}

\subsection{Non-genericity of the STP Contract}
Here we establish that the ``selling the project to the agent'' contract is generically violated by the agent's interim participation constraint. Formally,
\begin{remark} \label{nongeneric}
Identify a ($t$-action) decision problem as a point in Euclidean space $\mathbb{R}^{n \times t}$. For any decision problem with bounded payoffs $y \in \mathbb{R}^{n \times t}$ in which the principal optimally induces a non-degenerate distribution in the first-best benchmark and any neighborhood $U$ of $y$, there exists a decision problem $y' \in U$ for which the principal cannot attain efficiency via the STP contract.
\end{remark}
\begin{proof}
Suppose WLOG that in decision problem $y$, the principal can attain efficiency via a ``sell the product to the agent'' contract, and let $f_{\mathcal{H}}$ be the tangent hyperplane corresponding to optimal learning in the first-best benchmark. Let $a$ be an action that is taken with strictly positive probability by the principal in the first-best benchmark. Construct decision problem $y'$ by increasing the payoff from taking action $a$ in some state $l$ whose probability is nonzero at the belief at which the principal takes action $a$ in her optimal learning by $\varepsilon$. Evidently, the tangent hyperplane corresponding to optimal learning in the first-best benchmark for decision problem $y'$, $f_{\mathcal{H}}'$ is not equal to $f_{\mathcal{H}}$. Otherwise, the principal's payoff would be the same, which is a contradiction since her payoff at some on-path posterior has strictly increased.
\end{proof}

\subsection{Contracting on the Principal's Action}

As we discussed in the introduction of the main text, we frequently observe contracts in which an agent's remuneration is conditioned on the principal's action rather than the realized state. In this section we explore a simple two-state, two-action (for the principal) example with this feature. Our goal is to illustrate the following two things:
\begin{enumerate}[label={(\roman*)},noitemsep,topsep=0pt]
        \item In contrast to (the text's) Corollary 4.2, not all distributions over posteriors may be implementable; and
        \item Even if all distributions over posteriors are implementable, not all may be efficiently implementable (even with a risk-neutral agent and no limited liability).
    \end{enumerate}
Furthermore, note that even if a distribution can be implemented efficiently, the contract may distort the principal's behavior in her decision problem (relative to our main setting), introducing a new variety of inefficiency.

The principal's payoff from taking action \(a_1\) as a function of her posterior, \(x\), is \(\xi\left(1-x\right)\) (\(\xi \in \mathbb{R}_{++}\)), and her payoff from taking action \(a_2\) is \(\xi x\). As we noted in the text's introduction, a contract that conditions on the principal's action can successfully impel an agent to learn only if the principal has a secondary source of information. Therefore, we assume here that after the agent's report, the principal observes the realization of an exogenous signal before taking her action. We assume that the signal, \(\left(\pi,S\right)\), is binary, with \(p \coloneqq \pi\left(\left.h\right|\theta_{2}\right)\) and \(q \coloneqq \pi\left(\left.h\right|\theta_{1}\right)\) (and \(1 \geq p > q \geq 0\)).

Recall (from the text) our convention that $\gamma \coloneqq t_{2}^{1}$ is the transfer the agent obtains from sending message $x_L$ in state $\theta_2$, and $\beta \coloneqq t_1^2$ is the transfer from sending message $x_H$ in state $\theta_1$. We also define $\alpha \coloneqq t_{1}^{1}$ and $\delta \coloneqq t_2^2$. Given that the principal may not condition transfers on the state, these objects are now the expected action-contingent transfers from sending the two messages. We let \(\hat{a}_{2}^{1}\) denote the transfer paid to the agent if the principal takes action \(a_2\) following message \(x_L\); \(\hat{a}_{2}^{2}\), the transfer if the principal takes action \(a_2\) following message \(x_H\); and analogously for \(\hat{a}_{1}^{1}\) and \(\hat{a}_{1}^{2}\).

\begin{remark}
    Not all distributions over posteriors with support on \(\inter \Delta \left(\Theta\right)\) may be implementable.
\end{remark}
\begin{proof}
    Observe that for all sufficiently large \(\xi\), the differential payments to the agent are negligible for the principal in her decision problem. The principal's posterior following a report \(x_L\) and after observing \(h\) is \[\frac{x_L p}{x_L p + \left(1-x_L\right) q} < \frac{1}{2}\text{,}\] provided \(x_L\) is sufficiently low and \(p\) and \(q\) are sufficiently close. Likewise, the principal's posterior following a report \(x_H\) and after observing \(l\) is \[\frac{x_H \left(1-p\right)}{x_H \left(1-p\right) + \left(1-x_H\right) \left(1-q\right)} > \frac{1}{2}\text{,}\] provided \(x_H\) is sufficiently high and \(p\) and \(q\) are sufficiently close. Thus, \(\alpha = \beta = \hat{a}_{1}^{1}\) and \(\gamma = \delta = \hat{a}_{2}^{2}\); i.e., the agent's payoffs are state independent, so she will not report honestly.
\end{proof}
This example illustrates that poor secondary information on the part of the principal may impede information acquisition through an agent, as the principal is too tempted to ignore her own information in her decision problem. It is also clear that a contract in which the principal's signal is too informative in relation to what the agent is asked to acquire cannot be optimal. Consider, for instance, a binary signal (as in the remark's proof), but suppose that \(p\) and \(q\) are sufficiently close to \(1\) and \(0\), respectively, such that the principal's decision is independent of the agent's report. Evidently, the principal would prefer to have the agent acquire no information, which would leave the principal's decision unaffected, but save on the cost of paying the agent.

Accordingly, with a binary signal, the only pure-strategy candidate for a potential optimal arrangement is for the principal to ignore her private signal following one of the agent's reports, but to follow her private signal following the other. However, it may no longer be possible to leave the agent with no rents.
\begin{remark}
    Not all implementable distributions may be efficiently implementable (even with a risk-neutral agent and no limited liability).
\end{remark}
\begin{proof}
    We continue to assume that the principal acquires a binary signal, but now assume that \[\begin{split}
        &\max\left\{\frac{x_L \left(1-p\right)}{x_L \left(1-p\right) + \left(1-x_L\right) \left(1-p\right)q}, \frac{x_L p}{x_L p + \left(1-x_L\right) q}, \frac{x_H \left(1-p\right)}{x_H \left(1-p\right) + \left(1-x_H\right)\left(1-q\right)}\right\} < \frac{1}{2}\\ &< \frac{x_H p}{x_H p + \left(1-x_H\right)q}\text{,}
    \end{split}\]
    \textit{viz.,} the principal follows her signal after \(x_H\) but ignores it after \(x_L\). In this case, \(\alpha = \beta = \hat{a}_{1}^{1}\), \(\gamma = \left(1-q\right)\hat{a}_{1}^{2} + q \hat{a}^{2}_{2}\) and \(\delta = \left(1-p\right)\hat{a}_{1}^{2} + p \hat{a}^{2}_{2}\). As \(p \neq q\), there exists a family of solutions (recalling that \(A = \alpha - \gamma\) and \(B = \delta - \beta\) are pinned down by the distribution being implemented) with one degree of freedom. However, recall that efficient implementation requires that \(f_{\mathcal{H}}\) be tangent to \(v_0 - \kappa c\), which demands two degrees of freedom--that is, except in knife-edge cases, the agent, here, must be given rents.\end{proof}

\end{document}